\newif\iffinal
\newif\ifarxiv
\arxivtrue
\finaltrue

\documentclass[letterpaper,twocolumn,10pt]{article}
\usepackage[breaklinks,backref=page]{hyperref}
\usepackage{usenix}

\usepackage{hyperref}

\usepackage{microtype}
\usepackage{tikz}
\usepackage{subfig}
\usepackage{multirow}
\usepackage{booktabs}
\usepackage{graphicx}
\usepackage{xspace}
\usepackage{algorithm}
\usepackage[noend]{algpseudocode}
\usepackage{balance}
\usepackage{amsmath}
\usepackage{mathtools}
\usepackage{amsfonts}
\usepackage{amssymb}
\usepackage{amsthm}
\usepackage{makecell}
\usepackage{threeparttable}
\usepackage[available]{usenixbadges}

\usepackage[nosort]{cleveref} 

\newmuskip\normalthickmuskip
\newmuskip\normalthinmuskip
\AtBeginDocument{%
  \normalthickmuskip=\thickmuskip
  \normalthinmuskip=\thinmuskip
}
\everymath{%
  \thickmuskip=\muexpr\normalthickmuskip-(1\normalthinmuskip plus 1\normalthinmuskip)\relax
}
\everydisplay{\thickmuskip=\normalthickmuskip}
\let\texdisplaystyle\displaystyle
\renewcommand{\displaystyle}{\texdisplaystyle\the\everydisplay}

\algrenewcommand\algorithmicindent{0.5em}
\algrenewcommand\algorithmicrequire{\textbf{Input:}}
\algrenewcommand\algorithmicensure{\textbf{Output:}}

\theoremstyle{definition}
\newtheorem{definition}{Definition}[section]

\renewcommand{\thefootnote}{\arabic{footnote}}
\ifarxiv
\newcommand{\Sys}{MPK\xspace}
\newcommand{\sys}{MPK\xspace}
\else
\newcommand{\Sys}{TGX\xspace}
\newcommand{\sys}{TGX\xspace}
\fi

\newif\ifhighlight\highlightfalse
\newcommand{\rev}[1]{\ifhighlight\textcolor{blue}{#1}\else#1\fi} 

\newcommand{\commentout}[1]{}
\algnewcommand{\LeftComment}[1]{\Statex \(\triangleright\) #1}
\newcommand{\er}[1]{\mbox{\rm\em #1}}
\newcommand{\m}[1]{\mathcal{#1}}
\newcommand{\removed}[1]{}

\iffinal
\newcommand{\captionvspace}{-0.5em}
\else
\newcommand{\captionvspace}{-0.5em}
\fi

\newenvironment{revblock}{\ifhighlight\color{blue}\fi}{} 

\crefname{part}{\S}{\S\S}
\crefname{chapter}{\S}{\S\S}
\crefname{section}{\S}{\S\S}
\crefname{subsection}{\S}{\S\S}

\newcommand{\graph}{$t$Graph\xspace}
\newcommand{\graphs}{$t$Graphs\xspace}

\begin{document}

\title{\Sys: A Compiler and Runtime for Mega-Kernelizing Tensor Programs}
\author{
\rm{Xinhao Cheng}$^{1,*}$ \quad
\rm{Zhihao Zhang}$^{1,*}$ \quad
\rm{Yu Zhou}$^{1,*}$ \quad
\rm{Jianan Ji}$^{1,*}$ \quad
\rm{Jinchen Jiang}$^{2}$ \quad
\rm{Zepeng Zhao}$^{1}$ \\
\rm{Ziruo Xiao}$^{1}$ \quad
\rm{Zihao Ye}$^{3}$ \quad
\rm{Yingyi Huang}$^{3}$ \quad
\rm{Ruihang Lai}$^{1}$ \quad
\rm{Hongyi Jin}$^{1}$ \quad
\rm{Bohan Hou}$^{1}$ \\
\rm{Mengdi Wu}$^{1}$ \quad
\rm{Yixin Dong}$^{1}$ \quad
\rm{Anthony Yip}$^{1}$ \quad
\rm{Zihao Ye}$^{4}$ \quad
\rm{Songting Wang}$^{1}$ \quad
\rm{Wenqin Yang}$^{5}$ \\
\rm{Xupeng Miao}$^{6}$ \quad
\rm{Tianqi Chen}$^{1,3}$ \quad
\rm{Zhihao Jia}$^{1}$
\\[1em]
Carnegie Mellon University$^{1}$ \quad
Tsinghua University$^{2}$ \quad
NVIDIA$^{3}$ \\
University of Michigan$^{4}$ \quad
Independent Researcher$^{5}$ \quad
Peking University$^{6}$
}
\maketitle

{\let\thefootnote\relax\footnotetext{$^*$Equal contribution.}}

\begin{abstract}
We introduce Mirage Persistent Kernel (\sys), the first compiler and runtime system that automatically transforms multi-GPU model inference into a single high-performance mega-kernel. \Sys introduces an SM-level graph representation that captures data dependencies at the granularity of individual streaming multiprocessors (SMs), enabling cross-operator software pipelining, \rev{fine-grained overlap of computation and communication, and other optimizations that are infeasible under the conventional kernel-per-operator execution model}. The \sys compiler lowers tensor programs into optimized SM-level task graphs and generates fast CUDA implementations for each task, while the \sys in-kernel parallel runtime executes these tasks within a single persistent mega-kernel using decentralized scheduling across SMs. Together, these components provide end-to-end kernel fusion with minimal developer effort, while preserving the flexibility of existing programming models.
Our evaluation shows that \sys significantly outperforms existing kernel-per-operator LLM serving systems, achieving up to 1.7$\times$ lower end-to-end inference latency and pushing LLM inference performance close to the limits of the underlying hardware. \Sys is publicly available at \url{https://github.com/mirage-project/mirage}.
\end{abstract}

\section{Introduction}
\label{sec:intro}
Enabling high-performance inference of ML models on GPUs is critical for modern AI applications, since inference latency directly affects both user experience and serving cost. Today's ML systems generally express model computation as a tensor program structured as a directed acyclic graph, whose nodes denote tensor algebra operators (e.g., matrix multiplication) and whose edges represent tensors, i.e., the $n$-dimensional arrays produced and consumed by these operators.

Most existing systems execute each operator using a dedicated GPU kernel, either hand-optimized by domain experts~\cite{dao2023flash, ye2025flashinfer} or generated automatically by ML compilers~\cite{tillet2019triton, tvm, wu2025mirage}. However, this {\em kernel-per-operator} execution model limits several key cross-operator GPU optimizations.

First, modern GPUs impose an implicit {\em kernel barrier} between consecutive launches on the same stream to ensure that all threads from the previous kernel complete before any thread from the next kernel begins. While this mechanism correctly enforces data dependencies, it prevents {\em cross-operator software pipelining} and forces dependent operators to execute strictly sequentially. NVIDIA recently introduced {\em programmatic dependent launch} (PDL)~\cite{nvidia-pdl}, which allows partial overlap between kernels on the same stream. However, adopting PDL requires significant engineering effort, as it fundamentally alters kernel structure and control flow.

Second, the kernel-per-operator execution model prevents fine-grained compute-communication overlap. Since dependencies are represented only at the coarse granularity of operators, the runtime must enforce full-operator completion before launching dependent communication or computation. 
For example, when a matrix multiplication is followed by an all-reduce in separate kernels, the all-reduce must wait for the entire matrix multiplication to complete, even though each fragment of the all-reduce depends only on a subset of the multiplication output. Exploiting such opportunities requires representing and enforcing dependencies at a granularity finer than individual kernels.

Finally, kernel-per-operator execution may require launching hundreds to thousands of kernels for each inference iteration. To reduce launch overhead, current systems rely heavily on {\em CUDA Graphs}, which capture a sequence of GPU operations and replay them with low overhead. However, CUDA Graphs are largely static: any changes to control flow, tensor shapes, or data dependencies require re-instantiating or modifying the captured graph, limiting their flexibility for the dynamic workloads commonly seen in model inference.

A promising approach to overcoming these limitations is to fuse all computation and communication of model inference into a single {\em mega-kernel}, also known as a {\em persistent kernel}.
In this design, the system launches one GPU kernel to execute the entire model, including layer computations and inter-GPU communication, without interruption.

Mega-kernels address the limitations of kernel-per-operator execution in several ways.
First, they eliminate repeated kernel launch overhead by replacing many operator-level launches with a single kernel invocation.
Second, by fusing {\em all} operators into one kernel, they enable cross-operator software pipelining, allowing data for the next operator to be prefetched while computation for the current operator is still in progress.
Third, they support fine-grained overlap of computation and inter-GPU communication, enabling concurrent execution that more effectively hides communication latency.

Despite these benefits, automatically transforming an ML model into a high-performance mega-kernel remains challenging. Existing ML systems---such as PyTorch~\cite{pytorch}, Triton~\cite{tillet2019triton}, and TVM~\cite{tvm}---do not support end-to-end mega-kernel generation. 
Moreover, these systems rely on a fragmented ecosystem of specialized libraries: NCCL~\cite{nccl} or NVSHMEM~\cite{nvshmem} for communication, FlashInfer~\cite{ye2025flashinfer} or FlashAttention~\cite{dao2023flash} for attention, and CUDA or Triton for custom computation. 
This fragmentation makes it difficult to unify the entire inference pipeline within a single kernel.

We present {\em Mirage Persistent Kernel} (\sys), the first compiler and runtime system that automatically transforms multi-GPU model inference into a high-performance mega-kernel. 
\sys enables end-to-end kernel fusion with minimal developer effort: users can mega-kernelize a PyTorch model with only a few lines of code while achieving significant performance improvements compared to running the model in vanilla PyTorch with CUDA Graphs and {\tt torch.compile}. \sys combines the performance benefits of mega-kernels with the usability of existing ML frameworks.

\begin{figure}
    \centering
    \includegraphics[width=\linewidth]{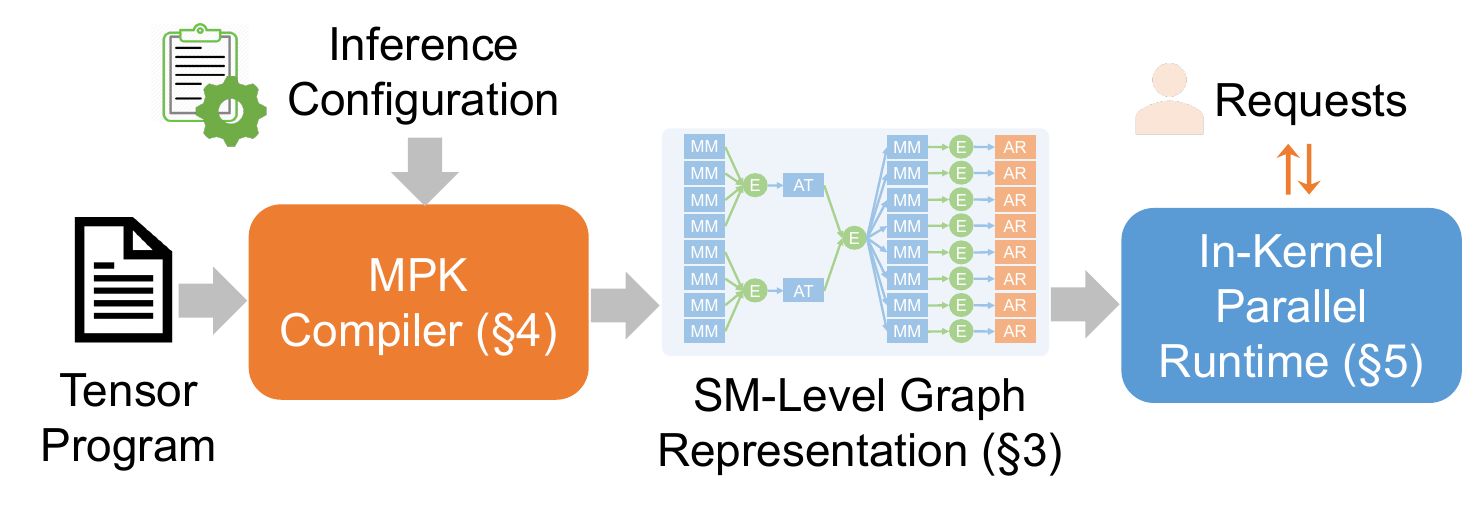}
    \vspace{\captionvspace}
    \caption{An overview of \sys.}
    \vspace{\captionvspace}
    \label{fig:overview}
\end{figure}

A key idea in \sys is to represent computation and inter-GPU communication at the granularity of individual streaming multiprocessors (SMs), rather than at the granularity of an entire GPU. 
\sys introduces an {\em SM-level graph representation}, called \graph, whose nodes denote {\em tasks} running on individual SMs and whose edges encode fine-grained dependencies between tasks. 
This representation exposes additional parallelism and enables optimizations such as cross-operator software pipelining and fine-grained kernel overlap, which are difficult to realize in conventional kernel-per-operator execution models.
\Sys realizes this idea using two key components shown in \Cref{fig:overview}.

%


\paragraph{The \Sys compiler.} 
The \Sys compiler takes a tensor program and an inference configuration as input and automatically transforms the program's computation graph into an optimized SM-level \graph tailored to the given inference configuration and GPU architecture.
The compiler applies a range of optimizations, including event fusion, graph normalization, and graph linearization, to reduce synchronization overhead and improve the performance of generated \graphs.
In addition, \Sys automatically generates fast CUDA implementations for individual tasks using existing superoptimization techniques~\cite{wu2025mirage}, ensuring efficient SM-level execution.

\paragraph{In-kernel parallel runtime.}
\Sys executes the SM-level \graph using an in-kernel parallel runtime embedded entirely within a mega-kernel, enabling fine-grained control over task execution and scheduling {\em without} additional kernel launches during model execution. 
To achieve this goal, the runtime partitions a GPU's SMs into {\em workers} and {\em schedulers}.
Each worker maintains a dedicated task queue and executes assigned tasks in a first-in-first-out order, while schedulers track dependencies across tasks and dispatch tasks once their prerequisites are satisfied. 
The \sys runtime uses an {\em event-driven, fully asynchronous} execution model to keep GPUs highly utilized. 
Finally, the runtime uses a {\em hybrid task-launch strategy} that combines just-in-time and ahead-of-time dispatch to minimize runtime overhead while preserving dynamic load balance across SMs.

\paragraph{Evaluation results.} 
We implement \sys as a PyTorch compiler backend: a PyTorch program can be compiled into an \sys mega-kernel with only a few lines of code changes.
We evaluate \sys on five widely used models across three generations of NVIDIA GPUs: A100, H100, and B200.
Even for workloads widely deployed and heavily optimized by existing kernel-per-operator systems, such as SGLang and vLLM for LLM serving, \sys outperforms current systems by 1.0--1.7$\times$ on both single- and multi-GPU deployments, pushing LLM inference performance close to hardware limits. 


\section{Background}
\label{sec:background}
This section first reviews the kernel-oriented GPU programming model and its limitations (\Cref{subsec:gpu-programming-model}), and then presents kernel fusion and mega-kernel techniques (\Cref{subsec:kernel_fusion}), which motivate the design of \sys.

\begin{figure}
    \centering
    \subfloat[Kernel barriers prevent cross-task pipelining.]{
    \includegraphics[scale=0.37]{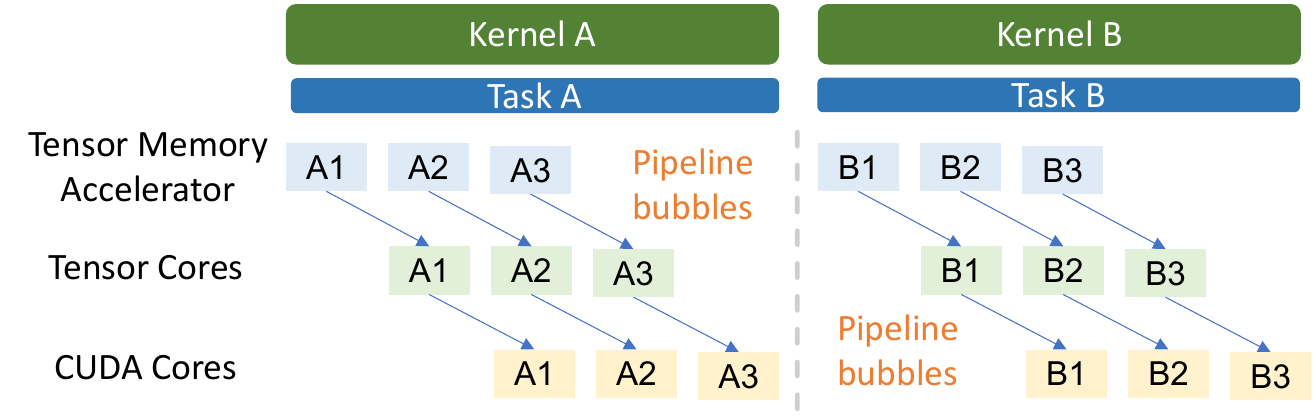}
    \label{fig:intra-task-pipelining}
    }
    \\
    \subfloat[\Sys enables both intra- and cross-task pipelining.]{
    \includegraphics[scale=0.37]{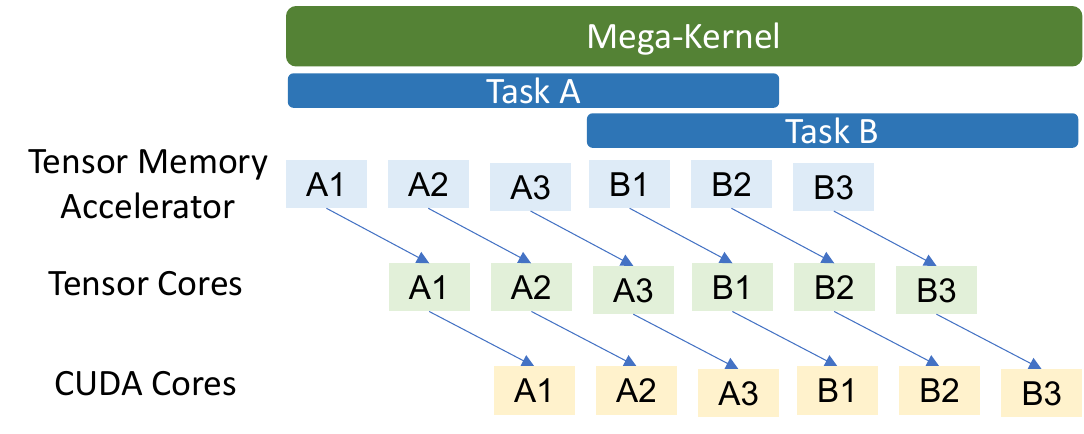}
    \label{fig:inter-task-pipelining}
    }
    \vspace{\captionvspace}
    \caption{Comparing how \sys and existing approaches support intra- and cross-task pipelining.}
    \vspace{\captionvspace}
    \label{fig:pipelining}
\end{figure}

\subsection{GPU Programming Model}
\label{subsec:gpu-programming-model}
On GPUs, computations are organized as {\em kernels}, each representing a function executed concurrently across many cores in a single-program, multiple-data (SPMD) fashion. A kernel consists of a grid of {\em thread blocks}, where each block is scheduled on a streaming multiprocessor (SM) and contains multiple {\em threads} that operate on individual data elements. Each thread has a private register file, while threads within the same block can cooperate through low-latency {\em shared memory} for data exchange and collective operations. All kernel inputs and outputs are stored in GPU {\em device memory}.

The conventional GPU programming model does not support direct synchronization across thread blocks within a kernel, \rev{because thread blocks are scheduled independently across SMs and may not all be resident simultaneously}. As a result, cross-operator dependencies  are enforced through {\em kernel barriers}, which are automatically inserted by the GPU runtime between consecutive kernels launched on the same stream.

While kernel barriers simplify dependency management, they also prevent key GPU optimizations such as cross-kernel software pipelining and fine-grained operator overlap.

\paragraph{Software pipelining.} GPU architectures are increasingly {\em heterogeneous}, integrating specialized accelerators such as tensor cores and tensor memory accelerators (TMAs). Since TMA load and store instructions execute asynchronously, data movement can proceed while tensor cores and CUDA cores perform computation. Fully exploiting these accelerators requires {\em software pipelining}---a technique that interleaves independent stages of computation and data movement across multiple iterations of tasks to maximize hardware utilization.

Existing systems implement {\em intra-task} pipelining, as shown in \Cref{fig:intra-task-pipelining}, where a single task is decomposed into multiple iterations. In this model, TMAs, tensor cores, and CUDA cores can simultaneously perform data transfer, matrix computation, and auxiliary operations for different iterations in a pipeline. However, kernel barriers restrict pipelining to within a single task, preventing {\em cross-task} pipelining and introducing pipeline bubbles that leave hardware resources underutilized.

\begin{figure}
    \centering
    \subfloat[Kernel barriers prevent overlapping {\tt MatMul} and {\tt AllGather}.]{
    \includegraphics[scale=0.45]{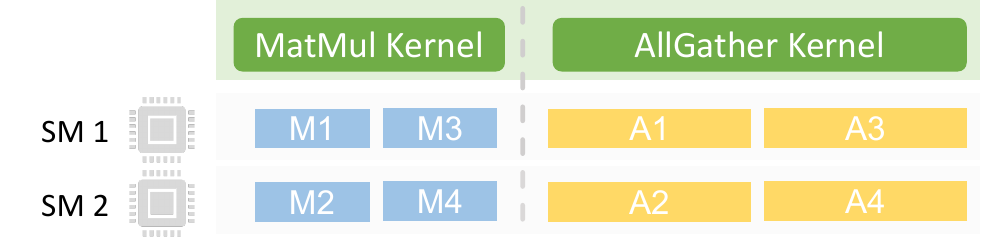}
    \label{fig:no-overlap}
    }
    \\
    \subfloat[\Sys enables fine-grained overlap of {\tt MatMul} and {\tt AllGather}.]{
    \includegraphics[scale=0.45]{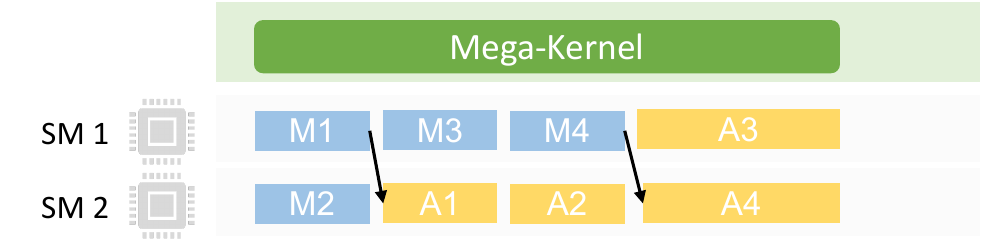}
    \label{fig:mpk-overlap}
    }
    \vspace{\captionvspace}
    \caption{Comparing how \sys and existing approaches support fine-grained kernel overlap between tasks. Data dependencies (black arrows in \Cref{fig:mpk-overlap}) ensure correctness.}
    \vspace{\captionvspace}
    \label{fig:overlap}
\end{figure}

\paragraph{Fine-grained kernel overlap.} Kernel barriers also preclude opportunities to overlap kernels that utilize different hardware resources (e.g., compute and communication), as they enforce dependencies at the granularity of entire kernels rather than individual data units. \Cref{fig:no-overlap} illustrates a common pattern in large language models (LLMs), where a {\tt MatMul} operator is followed by an {\tt AllGather} operator. Existing systems generally launch these as two separate kernels, requiring all thread blocks of the {\tt AllGather} kernel to wait until all thread blocks of the preceding {\tt MatMul} kernel complete.

In practice, the data dependency between {\tt MatMul} and {\tt AllGather} is much finer-grained: since {\tt AllGather} performs element-wise operations, each of its thread blocks only depends on the output of a single {\tt MatMul} thread block.
This dependency structure enables {\em fine-grained kernel overlap}, where different SMs can execute {\tt MatMul} and {\tt AllGather} in parallel, as long as fine-grained data dependencies are preserved. Such overlap allows the system to simultaneously utilize compute resources and communication bandwidth on modern GPUs, as identified in prior work~\cite{zhu2025nanoflow}. Achieving this overlap, however, requires proper synchronization between SMs at sub-kernel granularity, as shown in \Cref{fig:mpk-overlap}, which is not supported by conventional kernel barriers.

\begin{figure*}
    \centering
    \includegraphics[scale=0.36]{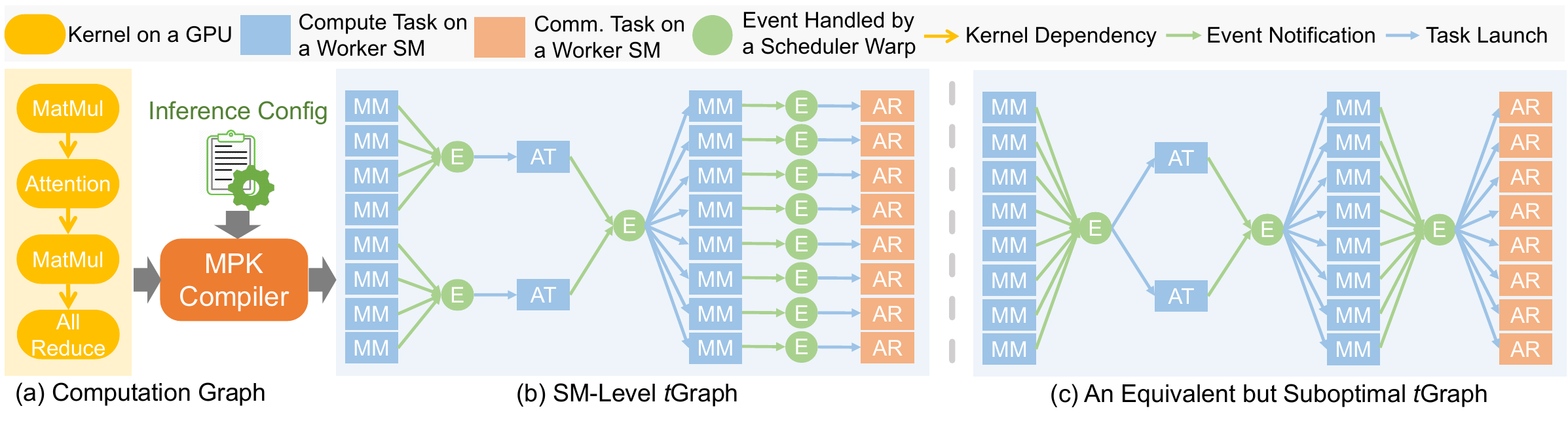}
    \vspace{\captionvspace}
    \caption{The \Sys compiler transforms a kernel-level computation graph into an optimized SM-level \graph. {\tt MM}, {\tt AT}, and {\tt AR} denote {\tt MatMul}, {\tt Attention}, and {\tt AllReduce} tasks, respectively.}
    \vspace{\captionvspace}
    \label{fig:task_graph}
\end{figure*}

\subsection{Kernel Fusion}
\label{subsec:kernel_fusion}
Kernel fusion eliminates kernel barriers by combining multiple GPU kernels that execute sequentially on the same data into a single, semantically equivalent kernel. Kernel fusion improves performance by avoiding \rev{materialization of} intermediate results, reducing device memory accesses, and eliminating kernel launch overheads.

Kernel fusion has been widely adopted in tensor program compilers. Frameworks such as PyTorch, JAX, and TVM employ rule-based heuristics to fuse adjacent kernels~\cite{pytorch, jax2018github, tvm}, while systems such as Mirage and TASO automatically discover fusion rules through compiler superoptimization~\cite{TASO, wu2025mirage}. However, existing compilers can only fuse small groups of local operators, as generating a single kernel that faithfully implements \rev{an entire complex tensor program} is computationally difficult and often infeasible.


The {\em mega-kernel} paradigm pushes kernel fusion to the extreme by fusing all computation and communication of a tensor program into one persistent kernel, using device-memory synchronization primitives to coordinate execution across SMs. 
Despite its performance benefits, current ML compilers such as PyTorch, Triton, JAX, and TVM do not support mega-kernel compilation. Existing mega-kernels are \rev{instead} handcrafted by GPU experts for specific models.
For example, FlashDMoE fuses mixture-of-experts computation and inter-GPU communication into a single kernel~\cite{aimuyo2025flashdmoe}, while Spector et al. manually designed and implemented a low-latency mega-kernel for LLAMA-1B~\cite{touvron2023llama,megakernel}. 

These manual approaches require substantial engineering effort and deep GPU expertise to mega-kernelize a tensor program. In contrast, \sys adopts a compiler-based approach that automatically transforms a tensor program into an optimized mega-kernel, eliminating the need for manual effort.

\section{SM-Level Graph Representation}
\label{sec:task_graph}

This section introduces {\em \graph}, a representation that expresses the computation of a tensor program at the granularity of individual streaming multiprocessors (SMs). \rev{Unlike conventional computation graphs, which expose dependencies only between tensor operators, \graph captures dependencies between SM-level units of work.} This fine-grained representation exposes additional parallelism and enables optimizations such as cross-operator software pipelining and fine-grained kernel overlap, both of which are not supported by the existing kernel-per-operator execution model.


\Cref{fig:task_graph} illustrates an example \graph, where each node represents either a {\em task} or an {\em event}. Each task---shown as a blue (or orange) rectangle---denotes a unit of computation (or communication) executed on a single SM. Each event---shown as a green circle---represents synchronization across tasks. Tasks and events alternate in the graph: every task only has outgoing edges to {\em triggering events} and incoming edges from {\em dependent events}. A task is ready for execution when its dependent events are all {\em activated} and notifies its triggering event upon completion. An event is activated once it has received notifications from all tasks associated with it.

This structure captures dependencies at a much finer granularity than traditional computation graphs. For example, multi-GPU LLM serving often involves a {\tt MatMul} operator followed by an {\tt AllReduce} operator (\Cref{fig:task_graph}a). Existing systems generally execute these operators sequentially because coarse-grained kernel barriers synchronize entire kernels. In contrast, SM-level task graphs can represent precise task-level dependencies: since {\tt AllReduce} performs \rev{element-wise communication and reduction}, each of its tasks depends only on one corresponding {\tt MatMul} task \rev{that produces its input tile}. By inserting fine-grained events between dependent task pairs, \sys can overlap compute-intensive {\tt MatMul} tasks with communication-intensive {\tt AllReduce} tasks, improving overall GPU utilization.

Multiple \graphs may represent the same computation graph. \Cref{fig:task_graph}c shows an alternative but suboptimal \graph where events capture only operator-level dependencies, analogous to traditional kernel barriers. \Cref{sec:compiler} describes how \sys generates high-performance task graphs by inferring {\em precise} data dependencies to maximize concurrency and minimize synchronization overheads.




\paragraph{Comparison with CUDA Graphs.} \graphs can be viewed as a lower-level extension of CUDA Graphs, sharing several structural similarities. Like CUDA Graphs, \graphs are statically instantiated and encode explicit dependencies among operations. However, while CUDA Graphs capture dependencies only at the kernel level, \graphs operate at the granularity of individual SM tasks and sub-kernel events. CUDA Graphs primarily describe kernel launch order and rely on stream semantics for synchronization, which confines overlap and fusion to kernel boundaries. In contrast, \graphs explicitly model both intra- and cross-operator dependencies, enabling fine-grained synchronization across SMs and overlap of computation and communication within a single kernel. This design allows \sys to exploit parallelism that is inaccessible to CUDA Graphs and other kernel-level execution models.
\begin{figure*}
    \centering
    \includegraphics[scale=0.39]{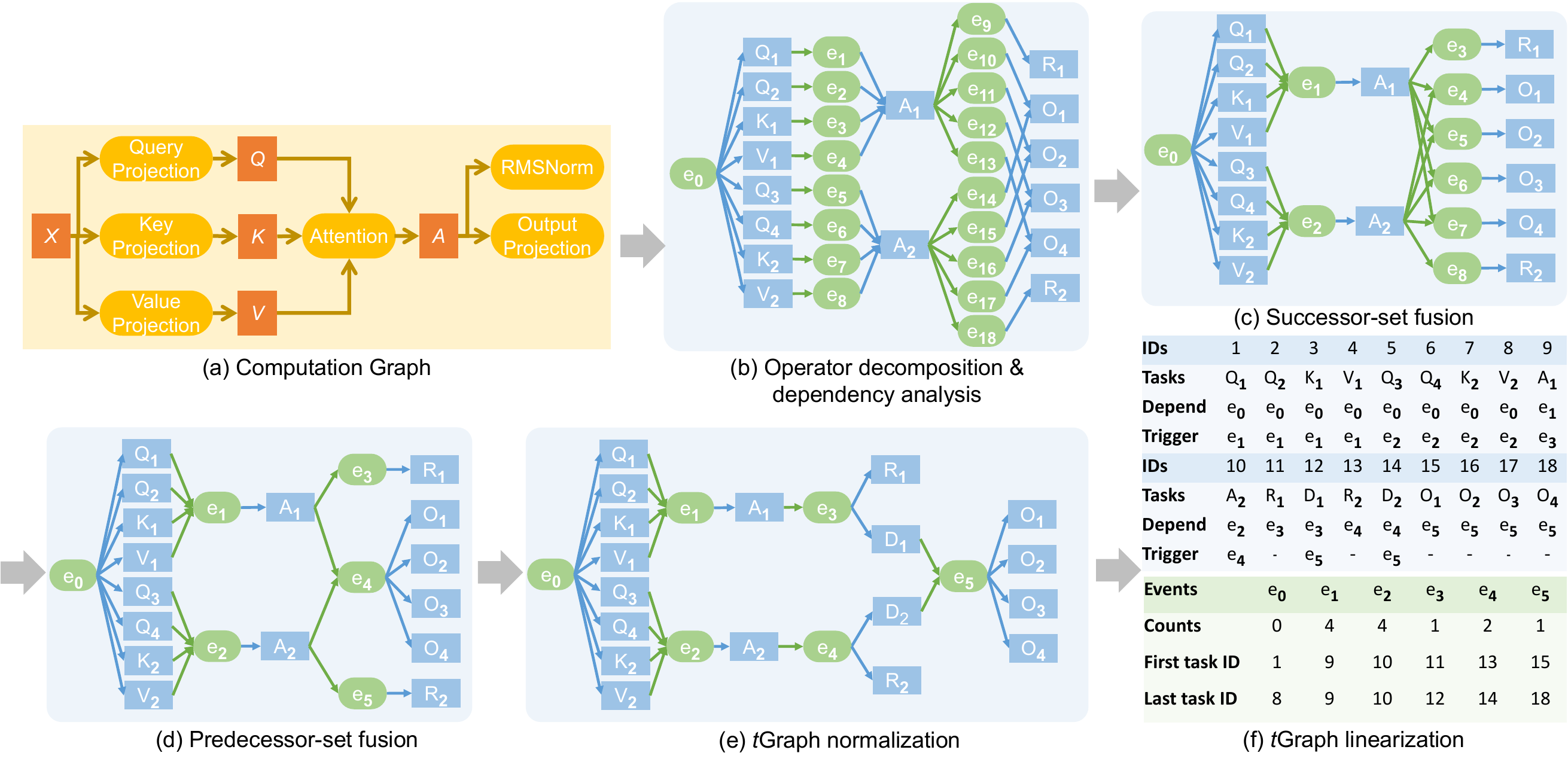}
    \vspace{\captionvspace}
    \caption{The \sys compiler workflow. In (b), $Q$, $K$, $V$, $A$, $O$, and $R$ denote the sets of tasks produced by decomposing the query projection, key projection, value projection, attention, output projection, and RMSNorm, respectively. $D_1$ and $D_2$ in (e) are dummy tasks inserted during \graph normalization to ensure that each task has a single triggering event. Finally, (f) shows how \Sys linearizes the \graph and stores the resulting structure, where tasks and events use a uniform, canonical representation.}
    \vspace{\captionvspace}
    \label{fig:compiler_workflow}
\end{figure*}

\section{The \Sys Compiler}
\label{sec:compiler}
This section presents the \sys compiler, which takes a computation graph and an associated inference configuration as input and generates an optimized \graph specialized for the target configuration and underlying GPU architecture. \Cref{fig:compiler_workflow} illustrates the end-to-end compilation workflow.

\subsection{\graph Generation}
\paragraph{Operator decomposition.}
The \sys compiler decomposes each operator in the input computation graph into a set of tasks by {\em partitioning} the operator's output tensors. Each task computes a {\em disjoint} subset of the output, allowing tasks from the same operator to execute in parallel across SMs. Most tensor algebra operators can be partitioned along multiple output dimensions; for example, the output tensor of a matrix multiplication can be tiled along both the row and column dimensions to expose parallelism.

The performance of a partitioning strategy depends on both the problem shape and the target GPU architecture. To discover an effective strategy, \sys selects a partitioning strategy that minimizes data loading from device memory to shared memory, since device memory accesses are significantly more expensive than shared-memory accesses or computation on CUDA cores and tensor cores. By default, \sys generates a number of tasks proportional to the number of SMs to promote load balance across SMs during execution. \Sys also provides an interface for users to specify custom partitioning strategies by setting the desired parallelization degree along each output dimension.

\paragraph{Dependency analysis.} \Sys uses {\em events} to capture dependencies between tasks. For any two operators sharing a tensor, \sys enumerates all pairs of tasks from the two operators and introduces an event $e$ for a task pair $(t_1, t_2)$ if and only if the output region produced by $t_1$ overlaps with the input region consumed by $t_2$. The event serves as a synchronization point indicating that $t_2$ cannot begin execution until $t_1$ has produced the required data. Accordingly, \sys inserts two edges, $(t_1, e)$ and $(e, t_2)$, into the resulting \graph. This fine-grained dependency analysis preserves all producer-consumer dependencies while exposing parallelism across independent tasks.

\paragraph{Event fusion.} 
\Sys applies two complementary forms of event fusion---{\em successor-set fusion} and {\em predecessor-set fusion}---to eliminate redundant synchronization points and simplify the constructed \graph. For an event $e$, we define two functions: $\er{InTasks}(e)$, the set of tasks that trigger $e$, and $\er{OutTasks}(e)$, the set of tasks that depend on $e$. These functions characterize when multiple events have identical dependency structure and can therefore be fused.

First, {\em successor-set fusion} merges events that serve as prerequisites for the same set of consumer tasks. Since these consumer tasks cannot begin execution until all such events are activated, representing the events separately provides no additional scheduling flexibility.

\begin{definition}[Successor-set fusion]
For any two events $e_1$ and $e_2$ in a \graph, successor-set fusion applies if and only if $\er{OutTasks}(e_1) = \er{OutTasks}(e_2)$. \Sys removes $e_1$ and $e_2$ from the \graph and introduces a fused event $e'$ with $\er{InTasks}(e') = \er{InTasks}(e_1) \cup \er{InTasks}(e_2)$ and $\er{OutTasks}(e') = \er{OutTasks}(e_1)$.
\end{definition}

For example, successor-set fusion merges events $e_{10}$ and $e_{14}$ in \Cref{fig:compiler_workflow}(b) into a new event, $e_4$ in \Cref{fig:compiler_workflow}(c), because both events are prerequisites for task $O_1$.

Second, {\em predecessor-set fusion} merges events that depend on the same set of producer tasks. Since such events are triggered simultaneously, maintaining them as separate synchronization nodes introduces unnecessary graph complexity.

\begin{definition}[Predecessor-set fusion]
For any two events $e_1$ and $e_2$ in a \graph, predecessor-set fusion applies if and only if $\er{InTasks}(e_1) = \er{InTasks}(e_2)$. \Sys removes $e_1$ and $e_2$ from the \graph and introduces a fused event $e'$ with $\er{InTasks}(e') = \er{InTasks}(e_1)$ and $\er{OutTasks}(e') = \er{OutTasks}(e_1) \cup \er{OutTasks}(e_2)$.
\end{definition}

For example, predecessor-set fusion merges events $e_4$, $e_5$, $e_6$, and $e_7$ in \Cref{fig:compiler_workflow}(c) into a single event, $e_4$ in the new \graph, because all four events depend on tasks $A_1$ and $A_2$.

A core challenge \Sys must address is representing dependencies between tasks and events efficiently. Because \Sys executes tasks and updates events in parallel across SMs, the runtime requires a {\em uniform} and {\em cheap} representation that avoids costly indirect indexing. Two challenges arise. First, a task may depend on and trigger an arbitrary number of events. A straightforward approach to representing tasks is reserving space for the maximum number of dependent and triggering events per task. However, this approach leads to significant memory overhead. Second, after event fusion, an event may trigger an arbitrary number of tasks. Representing these outgoing edges by allocating space for the maximum fan-out per event is also expensive.
\Sys addresses these challenges using two techniques: \graph normalization and \graph linearization.

\begin{figure}
    \centering
    \subfloat[A transformation reducing\\fan-out of a task to one.]{
    \includegraphics[scale=0.33]{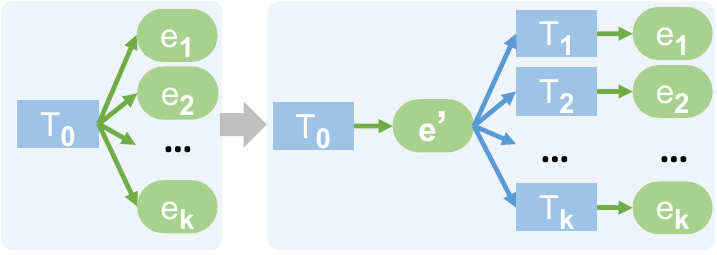}
    \label{fig:rewrite_1}
    }
    \subfloat[A transformation reducing\\fan-in of a task to one.]{
    \includegraphics[scale=0.33]{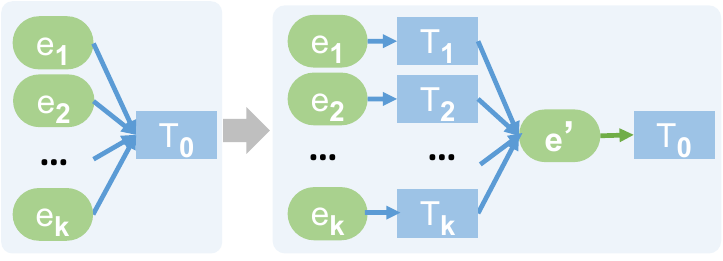}
    \label{fig:rewrite_2}
    }
    \vspace{\captionvspace}
    \caption{\Sys performs graph transformations to normalize an arbitrary \graph, ensuring that every task has at most one dependent event and at most one triggering event.}
    \vspace{\captionvspace}
    \label{fig:rewrites}
\end{figure}

\paragraph{\graph normalization.} 
\rev{Normalization bounds the dependency metadata stored for each task. When every task depends on and triggers at most one event, each task descriptor needs to store only one dependent-event identifier and one triggering-event identifier, rather than variable-length lists. \Sys achieves this property through the two rewrites shown in \Cref{fig:rewrites}, which transform an input \graph into a functionally equivalent form in which each task has at most one dependent event and at most one triggering event.} 


First, when a task $T_0$ triggers multiple events $e_1,\ldots,e_k$, \sys introduces a new event $e'$ and $k$ empty tasks $T_1,\ldots,T_k$, each of which performs no computation and depends on $e'$. After this transformation, $T_0$ triggers only $e'$, and each newly introduced task $T_i$ triggers exactly one of the original events $e_i$, as shown in \Cref{fig:rewrite_1}. 
This transformation ensures that every task has at most one {\em triggering} event. \Cref{fig:compiler_workflow}(e) shows how \sys applies this transformation to reduce the number of triggering events for $A_1$ and $A_2$ to one.


Second, when a task $T_0$ depends on multiple events $e_1,\ldots,e_k$, \sys introduces a new event $e'$ and $k$ empty tasks $T_1,\ldots,T_k$, each of which performs no computation and triggers $e'$. After this transformation, $T_0$ depends only on $e'$, and each newly introduced task $T_i$ depends on exactly one of the original events $e_i$, as shown in \Cref{fig:rewrite_2}. This transformation ensures that every task has at most one {\em dependent} event.


\graph normalization introduces additional tasks and events only when a \graph contains tasks with multiple fan-in or fan-out events. This situation typically arises when the original computation graph contains operators that can execute in parallel. For example, tasks $A_1$ and $A_2$ in \Cref{fig:compiler_workflow}(d) have two fan-out events because both the RMSNorm and output projection operators depend on attention and can therefore run in parallel. In practice, we observe negligible normalization overhead---always less than 1\% in our evaluation---because real-world models are usually ``deep'' (with many sequential operators), rather than ``wide'' (with many parallel operators).

\begin{algorithm}
\caption{\sys's \graph linearization algorithm. It is guaranteed that each task is enqueued into $T$ once and that each event is enqueued into $E$ once. Lines \ref{line:start}-\ref{line:end} ensure that all tasks depending on an event are consecutive in $T$.}
\label{algo:linearization}
\footnotesize
\begin{algorithmic}[1]
\Require{A normalized \graph $\mathcal{G}$}
\Ensure{A list of tasks $T$ such that for each event $e \in \mathcal{G}$, the tasks launched by $e$ are consecutive in $T$.}
\State $T \gets \varnothing$
\State $E \gets \{e \in \m{G}| e.{\rm counts}=0\}$ \Comment{Enqueue all events with no dependent tasks}
\While{$E$ is not empty}
\State $e \gets E.{\rm dequeue()}$
\ForAll{task $t \in \m{G}$}\label{line:start}
\If{$t.{\rm dependent\_event} = e$}
\State $T.{\rm enqueue}(t)$\label{line:end}
\State $e' \gets t.{\rm trigger\_event}$
\If{all tasks triggering $e'$ are in $T$}
\State $E.{\rm enqueue}(e')$
\EndIf
\EndIf
\EndFor
\EndWhile
\State
\State \Return $T$
\end{algorithmic}
\end{algorithm}

\paragraph{\graph linearization.} 
\rev{Linearization complements normalization: while normalization bounds the number of events associated with each task, linearization bounds the storage required to record the tasks associated with each event.} 
\graph normalization alone does not address the second representation challenge: after event fusion, an event may still need to trigger many tasks. For example, event $e_5$ in \Cref{fig:compiler_workflow}(e) triggers four tasks, which would otherwise require explicit storage for all outgoing task indices.

\Sys addresses this challenge using a breadth-first-search-based algorithm, shown in \Cref{algo:linearization}, to linearize a \graph. The algorithm assigns contiguous indices to all tasks triggered by the same event in the final task ordering. As a result, the fan-out of an event can be encoded compactly using only the first and last task indices, eliminating the need to store an explicit list of dependent tasks while preserving all dependencies.

\Cref{fig:compiler_workflow}(f) illustrates how \sys stores the linearized \graph in GPU device memory. For each task, \sys records only the indices of its dependent and triggering events. For each event, \sys stores the number of triggers required for activation; once activated, the runtime launches all tasks whose indices fall within the event's first and last task indices.

\subsection{Task Implementation Generation}
In addition to constructing a \graph, \sys must generate a device function for each task to execute on a GPU SM. \sys leverages prior work and uses a {\em compiler superoptimization} approach to automatically generate a high-performance implementation for each task.

Specifically, \sys performs superoptimization at the thread block level instead of the kernel level. Each compute task is associated with a reference PyTorch implementation, and \sys uses the Mirage superoptimizer~\cite{wu2025mirage} to search for an optimized thread-block graph, which is sent to the Mirage compiler to generate a CUDA implementation. The CUDA implementation includes intra-SM optimizations such as software pipelining, register reuse, and layout optimizations to reduce shared-memory bank conflicts.

\begin{figure}
    \centering
    \includegraphics[scale=0.4]{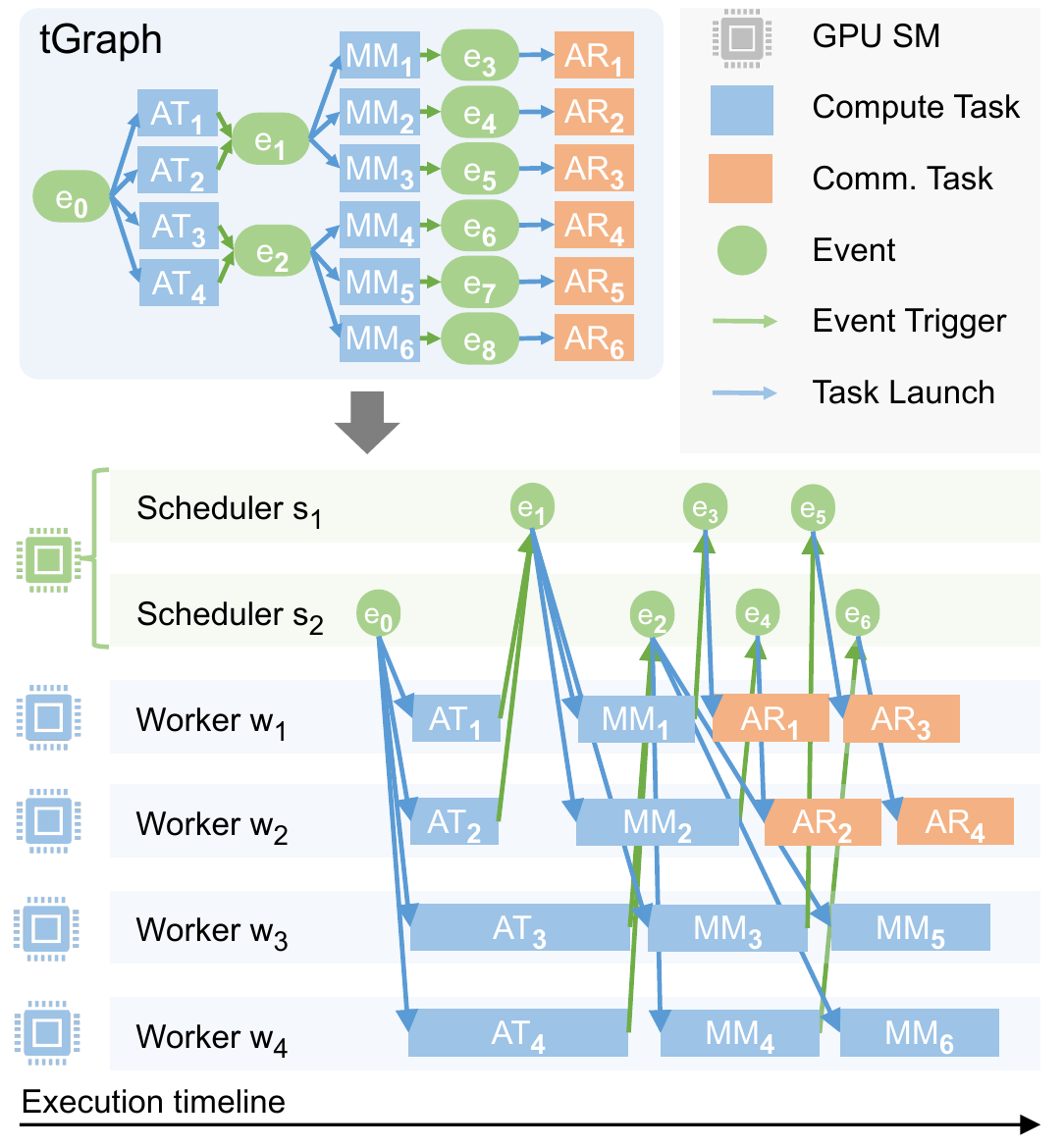}
    \vspace{\captionvspace}
    \caption{The \sys event-driven execution model. Circles denote events, and blue (or orange) rectangles denote compute (or communication) tasks, respectively. Edges from an event to a task correspond to task launches, while edges from a task to an event indicate that the task triggers the associated event upon completion. {\tt AT}, {\tt MM}, and {\tt AR} refer to attention, matrix multiplication, and AllReduce, respectively.}
    \vspace{\captionvspace}
    \label{fig:execution}
\end{figure}

\section{In-Kernel Parallel Runtime}
\label{sec:runtime}

\Sys employs an {\em in-kernel parallel runtime} that executes the \graph across all SMs within a single mega-kernel. This design eliminates repeated kernel launches and exposes fine-grained control over scheduling, synchronization, and execution order. Once launched, the mega-kernel continuously manages both computation and communication until the inference workload completes.

To support this execution model, \sys partitions a GPU’s SMs into {\em workers} and {\em schedulers}. Each worker runs on one physical SM and maintains an independent {\em task queue}. Workers execute a lightweight loop that repeatedly dequeues tasks, performs the associated computation or communication, and signals task completion by notifying the task's triggering event. This design ensures that workers are fully utilized while enabling asynchronous execution across operators.

Schedulers are organized at {\em warp} granularity, with each SM hosting four scheduler warps. Each scheduler maintains an event queue and repeatedly polls for newly activated events, dispatching the corresponding tasks to workers. The allocation of workers and schedulers is fixed at kernel-launch time and matches the GPU’s physical SM count, avoiding any dynamic role-switching overhead inside the kernel.

The remainder of this section details the in-kernel runtime architecture. \Cref{subsec:event_driven_execution} describes MPK’s event-driven execution model. \Cref{subsec:task_launch} introduces two complementary task-launch mechanisms, analyzes their trade-offs, and explains how \sys combines them to achieve low-latency and load-balanced execution. \Cref{subsec:runtime_optimizations} describes additional runtime optimizations that further reduce overhead and improve throughput.

\subsection{Event-Driven Execution}
\label{subsec:event_driven_execution}

\Sys executes a \graph using an {\em event-driven} model. Each \graph begins with a designated {\em start} event (e.g., $e_0$ in \Cref{fig:execution}), which has no prerequisites. This event is initially enqueued into a scheduler’s event queue. Upon dequeuing the event, the scheduler (e.g., $s_1$) launches all tasks that depend on it (e.g., $AT_1,\ldots,AT_4$). Each launched task is dispatched to a worker, which executes the task and, upon completion, notifies the triggering event associated with that task.

An event becomes \emph{activated} once all of its prerequisites have completed and thus have collectively triggered the event the required number of times. When an event is activated, it is enqueued into a scheduler’s event queue, allowing the runtime to continue propagating execution through the \graph. In this way, events serve as the mechanism for driving task execution, enabling fine-grained, asynchronous scheduling.

\rev{This event-driven model also allows \sys to adapt to the underlying
interconnect without requiring an explicit topology profile. Communication and computation are represented uniformly as tasks in the same \graph and dispatched by the same scheduler. A dependent compute task launches as soon as the communication tasks producing its inputs trigger the relevant event, regardless of the latency of the links they traverse. A faster link triggers its event sooner and releases dependent work earlier, while a slower link delays only the tasks that depend on it. The scheduler therefore does not require explicit knowledge of link bandwidths or topology distances. The same \graph and scheduler can thus handle heterogeneous interconnects, such as intra-node NVLink and inter-node networks, by reacting to data availability rather than relying on a static topology model.}

\begin{figure}
    \centering
    \subfloat[Just-in-time task launch.]{
    \includegraphics[scale=0.38]{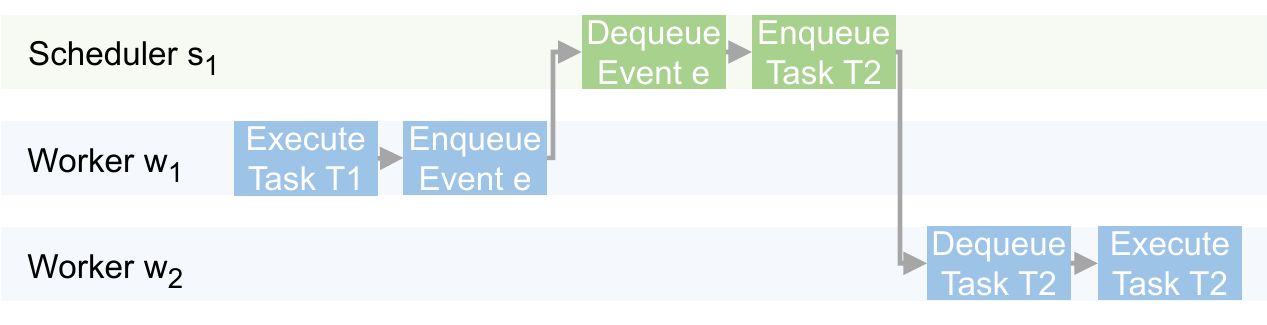}
    \label{fig:jit_launch}
    }
    \\
    \subfloat[Ahead-of-time task launch.]{
    \includegraphics[scale=0.38]{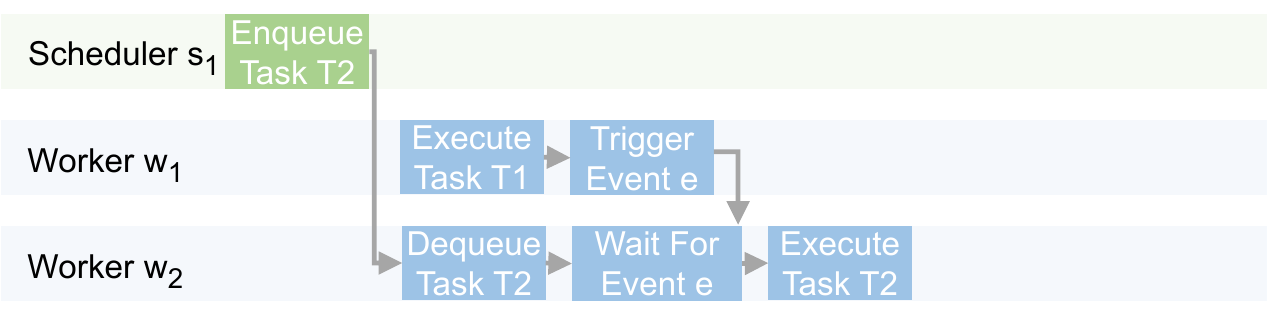}
    \label{fig:aot_launch}
    }
    \vspace{\captionvspace}
    \caption{Comparing JIT and AOT task launches.}
    \vspace{\captionvspace}
    \label{fig:aot_and_jit}
\end{figure}

\subsection{Hybrid Task Launch}
\label{subsec:task_launch}

A task can be enqueued into a worker's task queue either {\em just-in-time} (JIT) or {\em ahead-of-time} (AOT). In JIT mode, a scheduler assigns a task to a worker only after its dependent event has been fully activated; the task can therefore begin execution immediately after assignment. In AOT mode, the runtime pre-enqueues the task into a worker's task queue before its predecessor event is activated. The worker cannot execute the task until the dependent event is fully activated, and thus waits locally for the dependency to be satisfied.

JIT and AOT approaches provide complementary advantages. On the one hand, JIT launch allows \sys to adapt to workload imbalance. For example, attention in LLMs involves highly variable execution times due to data-dependent sequence lengths---requests with long sequence lengths take longer to finish than those with short ones. This variance makes static assignment ineffective. Under JIT launch, \sys launches downstream tasks (e.g., {\tt MatMul} or {\tt AllReduce}) only after the attention tasks that trigger them have completed. Workers that finish their attention tasks earlier can execute more downstream tasks, improving end-to-end latency and balancing load across SMs, as illustrated in \Cref{fig:execution}.


On the other hand, JIT launch involves higher latency due to additional worker–scheduler communication. \Cref{fig:aot_and_jit} illustrates the difference when launching task $T_2$ after event $e$ is triggered by task $T_1$. Under JIT launch (\Cref{fig:jit_launch}), the worker executing $T_1$ notifies a scheduler, which then dequeues event~$e$, launches $T_2$, and enqueues it into a worker’s task queue. The receiving worker must then dequeue $T_2$ before execution. This chain requires two synchronization steps (worker$\rightarrow$scheduler and scheduler$\rightarrow$worker). By contrast, under AOT launch (\Cref{fig:aot_launch}), $T_2$ has already been enqueued on a pre-assigned worker, which only needs to wait for event~$e$ to be activated. Thus, AOT launch requires only one synchronization step through the event trigger, reducing per-task launch latency.

\Sys uses {\em hybrid task launch} to combine the advantages of both approaches. During \graph construction, the compiler classifies each operator as JIT or AOT based on whether its execution time is data-dependent and likely to induce runtime imbalance. Operators with data-dependent durations (e.g., attention) are marked as JIT, and their downstream operators remain JIT until execution reaches a global barrier (i.e., an event that must be triggered by all upstream tasks). Such barriers eliminate accumulated imbalance, making subsequent operators suitable for AOT launch. All remaining operators are labeled AOT to minimize dispatch overhead. Labels apply at operator granularity: all tasks generated by the same operator share the same launch mode.

Workers maintain two queues, one for JIT tasks and one for AOT tasks. Workers always prioritize JIT tasks, as they are ready to execute immediately. When a worker exhausts its JIT queue, it checks whether the first AOT task’s dependent event has been fully activated; if so, the worker executes that AOT task. This design ensures that a worker blocks only when no ready work is available.

Schedulers handle only events that launch JIT tasks. All AOT tasks are {\em pre-enqueued} before execution begins. Because operators typically produce a number of tasks proportional to the number of workers (\Cref{sec:compiler}), \sys distributes AOT tasks across workers in a round-robin fashion to maintain balanced load. Pre-enqueuing AOT tasks reduces scheduler load and amortizes dispatch overhead, while JIT launch dynamically balances work across SMs.


\subsection{Runtime Optimizations}
\label{subsec:runtime_optimizations}
This subsection introduces runtime optimizations that minimize \sys's execution overhead.

\paragraph{Paged shared-memory abstraction.}
In conventional GPU programming models, shared memory is a fast on-chip memory private to each thread block. Shared memory exists only for the lifetime of a kernel: once the kernel finishes, its shared memory is automatically released. In the kernel-per-operator execution model, each kernel typically assumes {\em exclusive} access to the shared memory allocated to its thread block. However, this assumption prevents cross-task software pipelining (\Cref{sec:background}), which overlaps data loading for a subsequent task with computation in the current task, since both tasks may need shared memory at the same time.

To enable such pipelining, \Sys introduces a {\em paged shared-memory} abstraction. Shared memory is partitioned into fixed-size pages, and task implementations are modified to operate on pages instead of assuming a monolithic allocation. A task may {\em acquire} one or more pages based on its shared-memory footprint and must {\em release} the pages when they are no longer needed. Once a task releases any page, it is no longer permitted to acquire additional shared-memory pages, enforcing a monotonic usage pattern that simplifies scheduling. When the current task releases pages, \sys can immediately allocate available pages for the next task and begin data prefetching. This design enables fine-grained, on-demand allocation of shared memory within the mega-kernel execution model.

\paragraph{Cross-task software pipelining.}
To enable software pipelining across tasks executed on the same worker (\Cref{subsec:gpu-programming-model}), \sys decomposes each task into a {\em pre-loading} phase and a {\em compute} phase. The pre-loading phase issues data transfer instructions to fetch a chunk of the required tensor from device memory into shared memory, initializing the intra-task software pipeline without performing computation.

\Sys opportunistically overlaps the compute phase of the current task $T_1$ with the pre-loading phase of the subsequent task $T_2$ when two conditions hold: (1) $T_1$ has already issued all of its own data-transfer instructions, and (2) sufficient shared-memory pages are available for $T_2$'s pre-loading phase. This pipeline does not interfere with $T_1$'s execution because \sys inserts the necessary intra-SM synchronization barriers to ensure that $T_2$'s memory transfers do not conflict with ongoing data transfers for $T_1$.

\paragraph{Pre-fetching task descriptions.}
Each worker maintains both JIT and AOT task queues in device memory. Every task is associated with a task description that encodes its input tensors, output tensors, and configuration parameters; in our current implementation, each description occupies 352 bytes (\Cref{sec:impl}). To reduce enqueue/dequeue latency and hide device-memory access costs, \sys employs a lightweight prefetching mechanism that retrieves upcoming task descriptions into shared memory before they are needed.
\section{Evaluation}
\label{sec:eval}
This section evaluates \sys by answering four questions. First, \Cref{subsec:end_to_end} compares \sys's mega-kernel execution model with state-of-the-art kernel-per-operator systems. Second, \Cref{subsec:case_study_moe} studies how \sys supports dynamic workloads. Third, \Cref{subsec:multi_gpu} evaluates the scalability and efficiency of \sys for multi-GPU execution. Finally, \Cref{subsec:abalation} isolates the performance contributions of the main runtime optimizations in \sys.

We focus our evaluation on LLM serving for two reasons. First, LLM serving has several heavily optimized kernel-per-operator baselines, including SGLang and vLLM~\cite{zheng2024sglang, kwon2023vllm}; comparing against them provides a stringent benchmark that highlights the benefits of \sys's mega-kernel approach. Second, LLM serving naturally exhibits dynamic execution behavior, as different serving iterations may vary in batch size, sequence length, and the mixture of prefill and decode work. This variability creates heterogeneous workloads that stress both the compiler and the runtime. Although our evaluation focuses on LLM serving, the \sys compiler and runtime are model-agnostic and can support general DNN architectures.

\subsection{Implementation Details}
\label{sec:impl}

We implement \sys as a PyTorch compiler backend. A PyTorch program can be compiled into an \sys mega-kernel via PyTorch’s compilation interface by specifying \sys as the backend, i.e., {\tt torch.compile(backend=\sys)}. This call invokes the \sys compiler, which generates a mega-kernel and returns it as a callable PyTorch function. Invoking this function issues a single launch of the generated mega-kernel.

The current \sys implementation consists of approximately 44K lines of C++, 42K lines of CUDA, and 10K lines of Python. The in-kernel parallel runtime is written in CUDA and uses semaphores in device memory to coordinate workers and schedulers. The \sys compiler, implemented in C++ and Python, automatically transforms an input tensor program into an optimized \graph tailored to specific GPU types. For compute tasks, the compiler integrates the Mirage superoptimizer~\cite{wu2025mirage} to automatically generate optimized CUDA implementations and uses NVSHMEM~\cite{nvshmem} to support in-kernel inter-GPU communication.


Our implementation includes several optimizations to minimize runtime overhead and support dynamic workloads.

\paragraph{Task-launch overhead.}
Because \sys decomposes computation into tasks that are substantially finer-grained than traditional GPU kernels, minimizing per-task launch overhead is essential for performance. \sys uses several techniques to keep task-launch costs low. First, the runtime uses lightweight workers and schedulers: event queues and task queues are implemented as circular buffers in GPU device memory, and enqueue and dequeue operations rely only on low-cost {\tt atomicAdd} instructions. Second, \sys uses decentralized scheduling, in which each scheduler assigns tasks using local state. This design avoids global coordination and eliminates the communication and synchronization overheads inherent to globally coordinated scheduling.

Although the current implementation uses decentralized scheduling, the runtime is designed to support alternative policies, including globally coordinated scheduling, with minor code changes. Exploring these policies and their performance trade-offs is an interesting direction for future work.


\paragraph{Supporting runtime dynamism.}
To demonstrate \sys's ability to support highly dynamic workloads, we extend the system with mechanisms required for LLM serving, including continuous batching~\cite{yu2022orca} and paged attention~\cite{kwon2023vllm}. When processing the start event of a \graph, the scheduler prepares a new decoding iteration by (1) removing completed requests from the previous iteration, (2) admitting newly arrived requests, and (3) updating per-request KV-cache metadata. This logic executes as a single task, and the KV-cache metadata is stored in device memory for direct access by attention tasks.

To handle the dynamic batch sizes intrinsic to LLM serving, \sys generates multiple \graphs specialized for representative batch sizes, using powers of two up to the maximum batch size. At runtime, the scheduler selects the appropriate \graph based on the current batch size. This approach allows the compiler to generate \graphs optimized for specific batch sizes while preserving flexibility for dynamic serving workloads.

\begin{table}
    \centering
    \caption{\sys configuration in our evaluation.}
    \small
    \label{tab:benchmarks}
    \begin{tabular}{l|rrr}
    \toprule
    {\bf GPU} & {\bf \# SMs} & {\bf \# Workers} & {\bf \# Schedulers}\\
    \midrule
    A100 & 108 & 104 & 16 \\
    H100 & 132 & 128 & 16 \\
    B200 & 148 & 144 & 16 \\
    \bottomrule
    \end{tabular}
    \vspace{\captionvspace}
\end{table}

\subsection{Experimental Setup}
\begin{figure*}
    \centering
    \vspace{\captionvspace}
    \includegraphics[width=\linewidth]{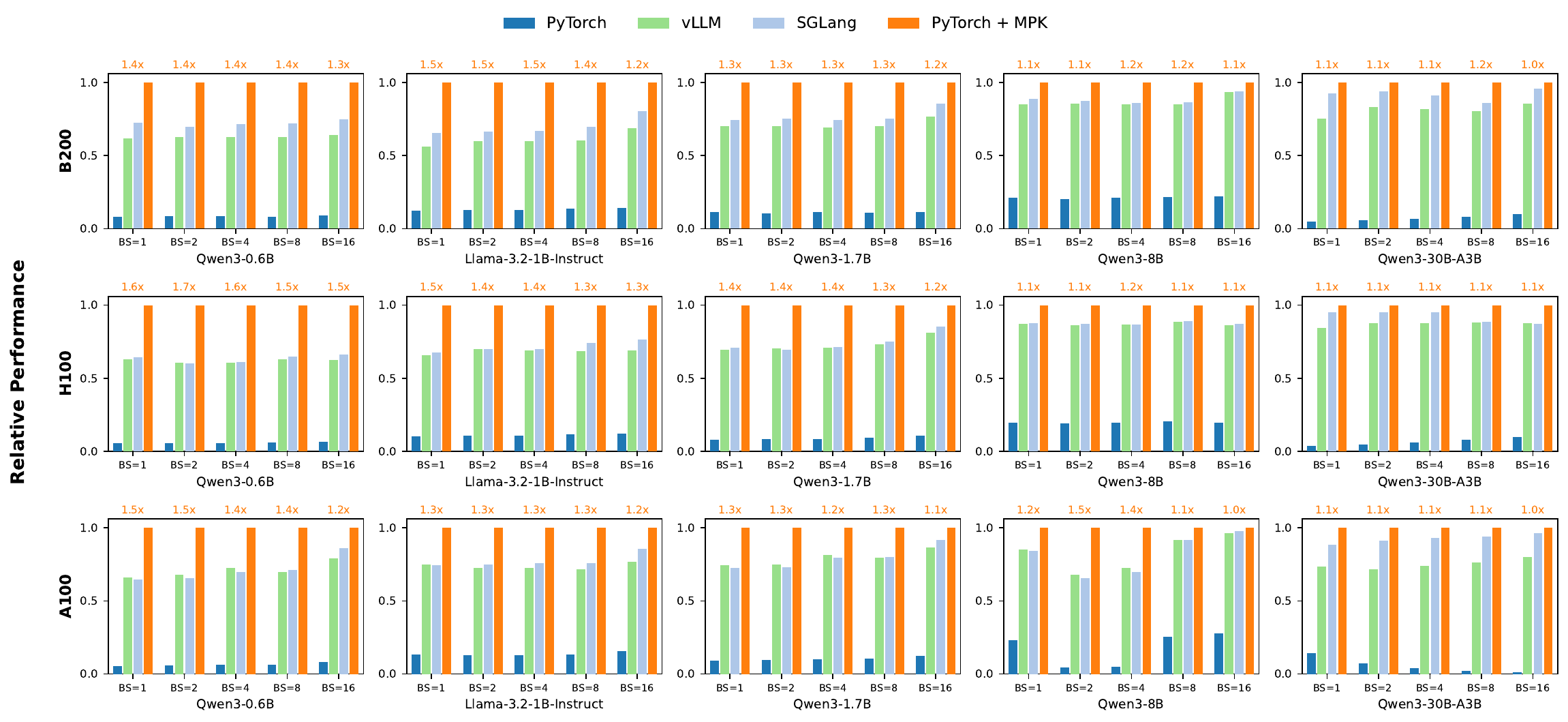}
    \vspace{-1em}
    \caption{End-to-end comparison of \sys with existing systems across five models on A100, H100, and B200 GPUs. All results are normalized to \sys; higher is better. The value above each \sys bar reports its speedup over the best-performing baseline.}
    \vspace{\captionvspace}
    \label{fig:end_to_end}
\end{figure*}

We evaluate \sys on five widely deployed LLMs that span different parameter scales and architectural families, and on three generations of NVIDIA GPUs: A100, H100, and B200.
\Cref{tab:benchmarks} summarizes the \sys configuration for each GPU. In all experiments, \sys reserves four SMs for schedulers, allocating 16 scheduler warps in total because each SM can host up to four active scheduler warps. The remaining SMs are used as workers. We set the shared-memory page size to 32~KB on all GPUs, which yields 5, 7, and 7 shared-memory pages per SM on A100, H100, and B200, respectively.
\Cref{fig:end_to_end} shows the evaluated LLMs, which include both dense and mixture-of-experts models across multiple model sizes.

To control for variability in request-arrival patterns, all experiments are conducted in an offline batched-inference setting while varying the maximum batch size. Each request uses a prompt length of 64 tokens and generates 1024 output tokens. This methodology eliminates server-side stalls caused by insufficient request concurrency and enables a controlled comparison of system-level performance.



\subsection{End-to-end Results}
\label{subsec:end_to_end}

We first compare the end-to-end serving performance of \sys with SGLang and vLLM, two state-of-the-art LLM serving systems. Both SGLang and vLLM use the kernel-per-operator approach and rely on specialized kernel libraries, including FlashInfer~\cite{ye2025flashinfer} and FlashAttention~\cite{triton-flashattention} for attention, cuBLAS and cuTLASS~\cite{cutlass} for matrix multiplication, and CUDA or Triton~\cite{tillet2019triton} for other operators. All systems load model architectures from HuggingFace Transformers~\cite{huggingfacemodels}, use bfloat16 precision, and enable paged attention~\cite{kwon2023vllm} and continuous batching~\cite{yu2022orca}. The key architectural difference is that \sys integrates page allocation and request scheduling directly into the mega-kernel. In contrast, SGLang and vLLM perform these operations on the CPU, incurring additional host–device synchronization and dispatch overheads.

For each model, we evaluate all three systems on B200, H100, and A100 GPUs with maximum batch sizes from 1 to 16, and report serving throughput. 


\Cref{fig:end_to_end} shows the end-to-end throughput results. For single-batch inference, \sys improves serving performance by 1.0–1.7$\times$ across models and hardware. The improvements are most significant for smaller models and newer GPU generations. This trend is expected because three overheads become more significant as the amount of computation per token decreases and GPU hardware becomes faster:
(1) kernel-per-operator approaches incur kernel-switch overheads, even when using CUDA Graphs; (2) kernel boundaries introduce pipeline bubbles and prevent cross-task pipelining (\Cref{fig:pipelining}); and
(3) SGLang and vLLM perform page allocation and request scheduling on the CPU, adding CPU–GPU synchronization delays.
\Cref{subsec:abalation} quantifies the impact of these optimizations.

These results show that \sys is well-suited for latency-sensitive serving scenarios, such as single-batch decoding, where reducing per-token latency is critical. For example, on Qwen3-8B running on an A100 GPU, \sys reduces per-token decoding latency from 14.5~ms, achieved by highly optimized systems such as vLLM and SGLang, to 12.5~ms. This approaches the approximate hardware lower bound of 10~ms, estimated by loading 16~GB of model parameters at 1.6~TB/s memory bandwidth.

Beyond performance, \sys also improves programmability. vLLM and SGLang require substantial engineering effort to optimize new models and integrate specialized kernels. In contrast, \sys takes a compiler-based approach that automatically mega-kernelizes a PyTorch model with only a few lines of code changes. As a result, \sys combines high performance with a familiar PyTorch development workflow, achieving more than 10$\times$ speedup over native PyTorch.

\begin{figure}
    \centering
    \includegraphics[width=0.8\linewidth]{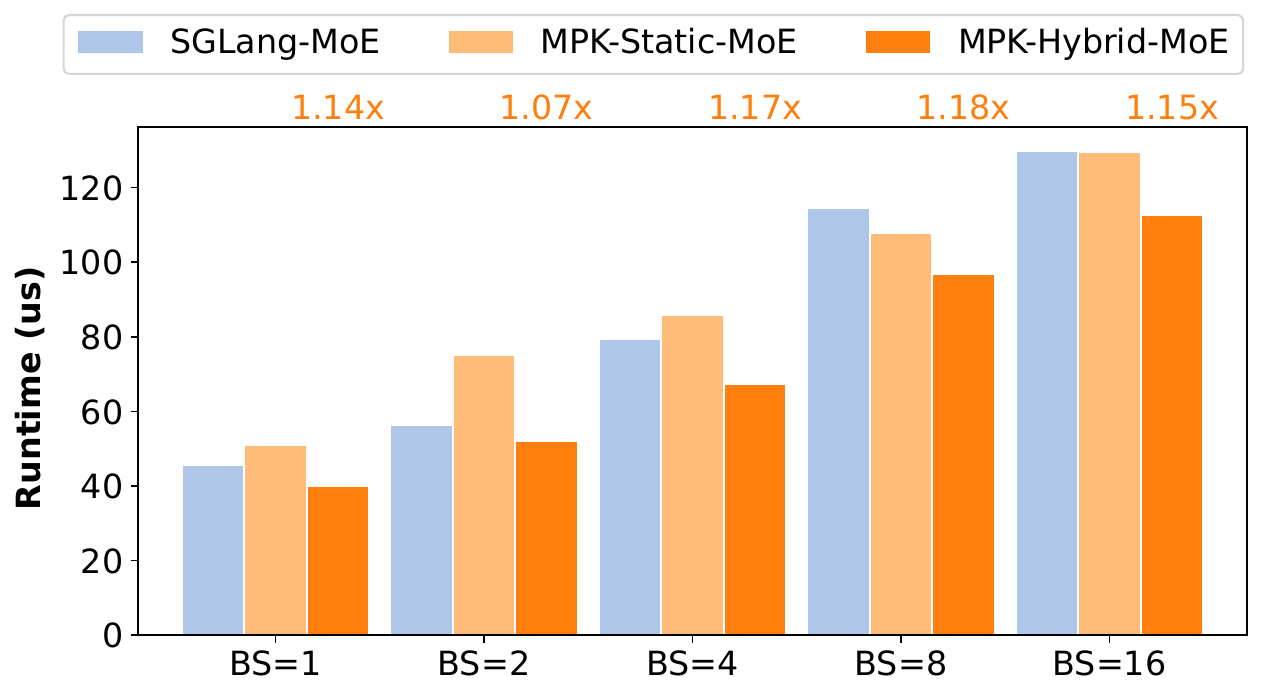}
    \vspace{\captionvspace}
    \caption{Comparing \sys with existing systems for Qwen3-30B-A3B on B200. Each value represents the actual MoE runtime in microseconds for each approach (lower is better), and the numbers above the bars indicate the speedup achieved by \sys-Hybrid-MoE over SGLang-MoE.}
    \vspace{\captionvspace}
    \label{fig:moe_ablation}
\end{figure}

\subsection{Case Study: Mixture-of-Experts}
\label{subsec:case_study_moe}
To efficiently serve dynamic workloads such as mixture-of-experts (MoE) models, \sys implements two MoE-specific optimizations: a hybrid workload balancer and a fused gather–GEMM implementation.

\begin{revblock}\paragraph{Representing expert parallelism.}
\Sys represents expert parallelism using the same task abstraction  used for other operators. An MoE block is lowered into a chain of SM-level tasks in the \graph (\Cref{sec:compiler}), including routing, dispatch, expert computation, and combine. The dispatch and combine stages capture the all-to-all communication required by expert parallelism, and \sys lowers them into inter-GPU data-transfer tasks in the same way it lowers other collective operations (\Cref{subsec:multi_gpu}). Because communication tasks and expert-computation tasks reside in the same \graph and are dispatched by the same event-driven scheduler, the runtime overlaps all-to-all communication with computation as soon as the relevant events are triggered, without relying on CUDA streams or host-side synchronization. The two MoE-specific optimizations described below operate within this task representation.\end{revblock}

\paragraph{Hybrid workload balancer.}
Because the number of tokens routed to each expert is known only at runtime, choosing an effective workload partition statically is challenging. A naive static strategy assigns a fixed group of SMs to preassigned experts. However, under skewed routing distributions, this strategy can lead to severe load imbalance: some SM groups become oversubscribed while others remain underutilized. At the other extreme, a fully dynamic strategy based on persistent Grouped-GEMM~\cite{moegroupedgemm} can balance work across SMs, but introduces fine-grained synchronization overheads.

\sys therefore uses a {\em hybrid} strategy that combines static structure with runtime adaptivity. At compile time, the compiler partitions the MoE computation into expert-specific tasks. At runtime, each task receives a meta-tensor produced by {\tt topk–softmax} containing global MoE information, including the number of activated experts and tokens assigned to each expert. Using this metadata, tasks refine their workload allocation dynamically and split work more evenly while avoiding the overheads of fully dynamic scheduling. As shown in \Cref{fig:moe_ablation}, this hybrid strategy consistently outperforms purely static partitioning across all  batch sizes.

\paragraph{Fused gather-GEMM.} 
To use tensor memory accelerators (TMAs) on Hopper and Blackwell GPUs, conventional MoE implementations first gather tokens routed to the same expert into a contiguous memory layout. For Qwen3-30B-A3B at batch size one on SGLang, this preprocessing step accounts for up to 11\% of total MoE execution time. In \sys, implementing this step as a separate preprocessing task would also introduce additional scheduling overhead.

\sys addresses this issue by replacing the TMA-based gather with an asynchronous token-level copy integrated directly into the data-loading phase of the GEMM tasks. This fusion eliminates the standalone gather kernel and avoids additional scheduling points while preserving efficient memory movement. As a result, \sys with fused gather-GEMM achieves consistent speedups over SGLang's implementation.

\begin{figure}
    \centering
    \vspace{\captionvspace}
    \includegraphics[width=\linewidth]{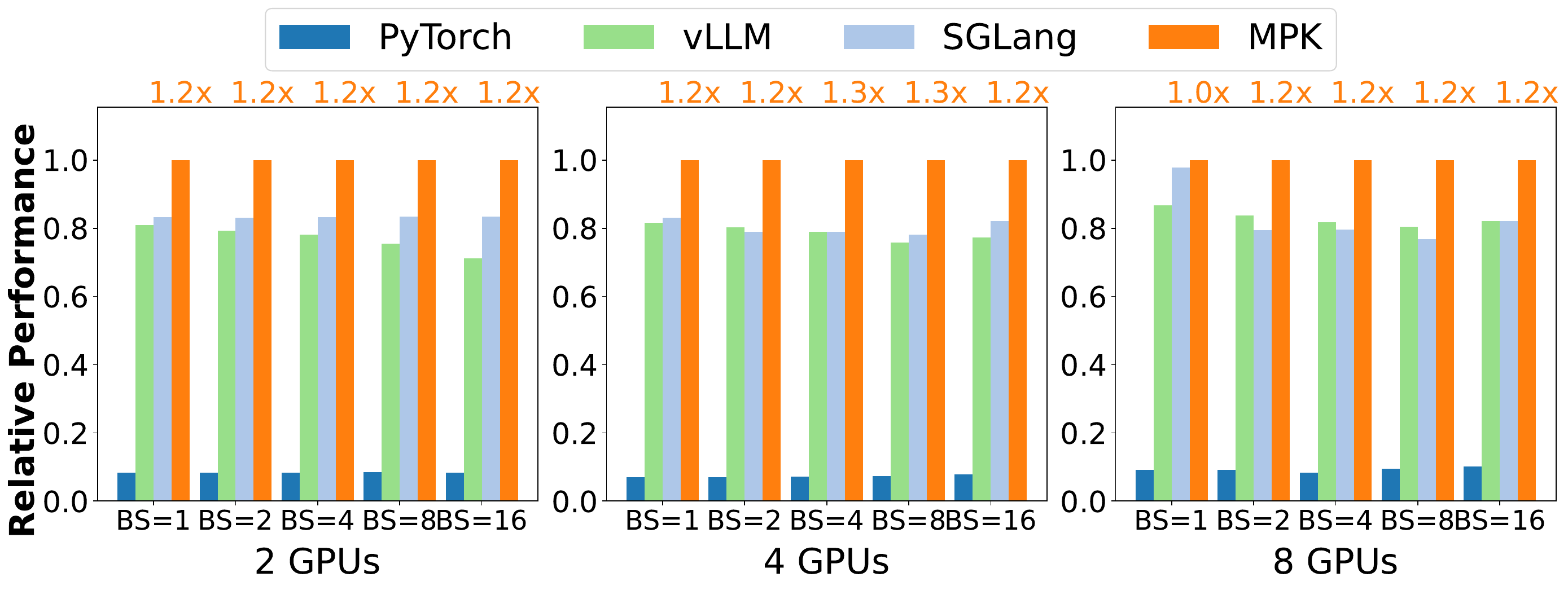}
    \vspace{-1em}
    \caption{\rev{Multi-GPU comparison of \sys with existing systems for Qwen3-1.7B on H100 GPUs using tensor parallelism. All results are normalized to \sys; higher is better.}}
    \vspace{\captionvspace}
    \label{fig:end_to_end_multigpu}
\end{figure}

\subsection{Multi-GPU Results}
\label{subsec:multi_gpu}

We evaluate the scalability of \sys across multiple GPUs on an NVIDIA H100 DGX instance. As in the baseline systems, we use tensor model parallelism, following Megatron-LM~\cite{Megatron}. Users specify tensor-parallel execution by inserting {\tt AllReduce} after attention and gated MLP blocks. \sys then automatically compiles these collective operators into two types of tasks: inter-GPU data-transfer tasks, implemented using NVSHMEM's {\tt nvshmem\_signal\_wait\_until}, and local reduction tasks. This decomposition converts synchronous collective communication into asynchronous tasks that can be integrated into \sys's task-based, event-driven runtime.

\Cref{fig:end_to_end_multigpu} shows the multi-GPU results. Compared with PyTorch, which uses a combination of hand-optimized kernels, CUDA Graphs, and {\tt torch.compile}, MPK’s mega-kernel execution improves throughput by up to 10$\times$. Compared with highly optimized serving systems such as SGLang and vLLM, \sys achieves 1.1--1.4$\times$ speedups when scaling to 8 H100 GPUs. These gains come from three optimizations missing in kernel-per-operator baselines: (1) \sys integrates page allocation and request scheduling directly into the mega-kernel, eliminating CPU-side dispatch overheads; (2) MPK’s asynchronous execution model overlaps compute tasks with collective communication; and (3) \sys eliminates kernel barriers and enables cross-task software pipelining. \Cref{subsec:abalation} analyzes the latter two optimizations in detail.


\begin{figure}
    \centering
    \vspace{\captionvspace}
    \includegraphics[width=0.8\linewidth]{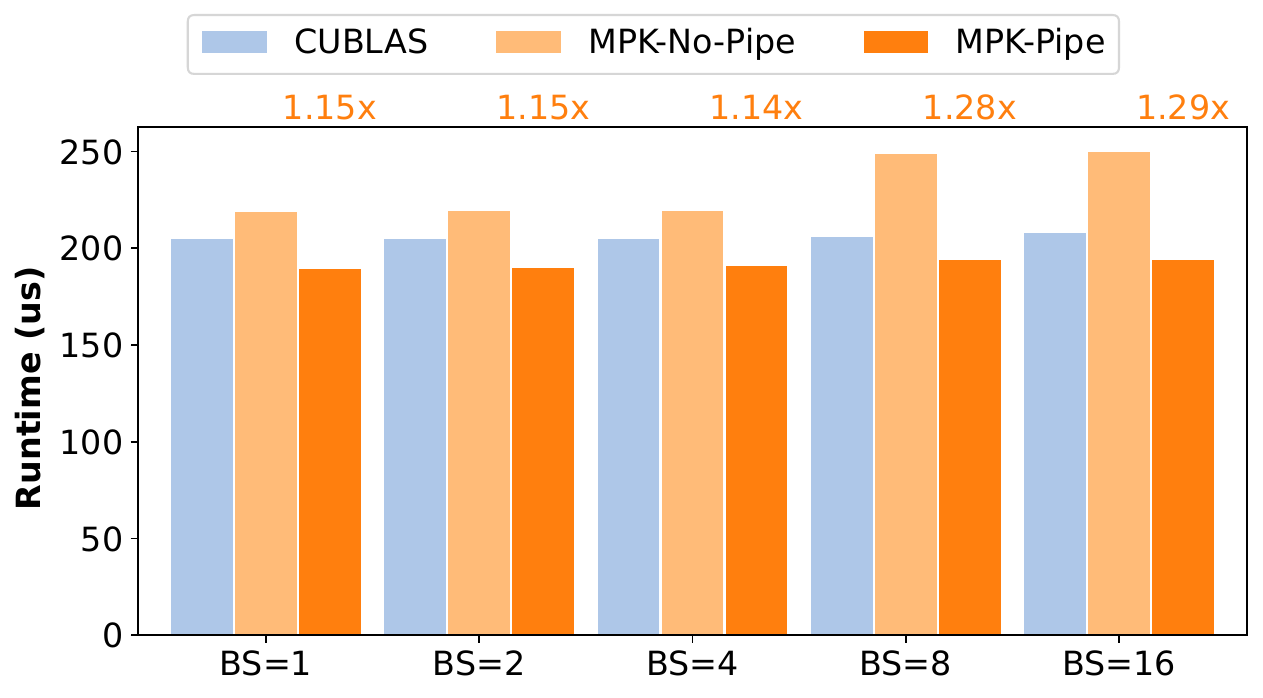}
    \vspace{\captionvspace}
    \caption{Ablation study of cross-task pipelining. We measure the runtime of the final linear layer in Qwen3-8B on an NVIDIA B200 GPU and report execution time in microseconds; lower is better. The value above each bar reports the speedup of \sys-Pipe over \sys-No-Pipe.}
    \vspace{\captionvspace}
    \label{fig:pipe_ablation}
\end{figure}

\subsection{Ablation Study}
\label{subsec:abalation}
This section evaluates the impact of three runtime optimizations enabled by \sys: cross-task pipelining, compute--communication overlap, and kernel-launch reduction.

\paragraph{Cross-task pipelining.} As described in \Cref{subsec:runtime_optimizations}, \sys enables cross-task pipelining by preloading chunks of input tensors for the next task while the current task is executing. \Cref{fig:pipe_ablation} evaluates the impact of this optimization on the final linear layer in Qwen3-8B. Cross-task pipelining reduces task runtime by 1.2--1.3$\times$ and even outperforms cuBLAS-based compiled kernels.

\begin{figure}
    \centering
    \includegraphics[width=0.8\linewidth]{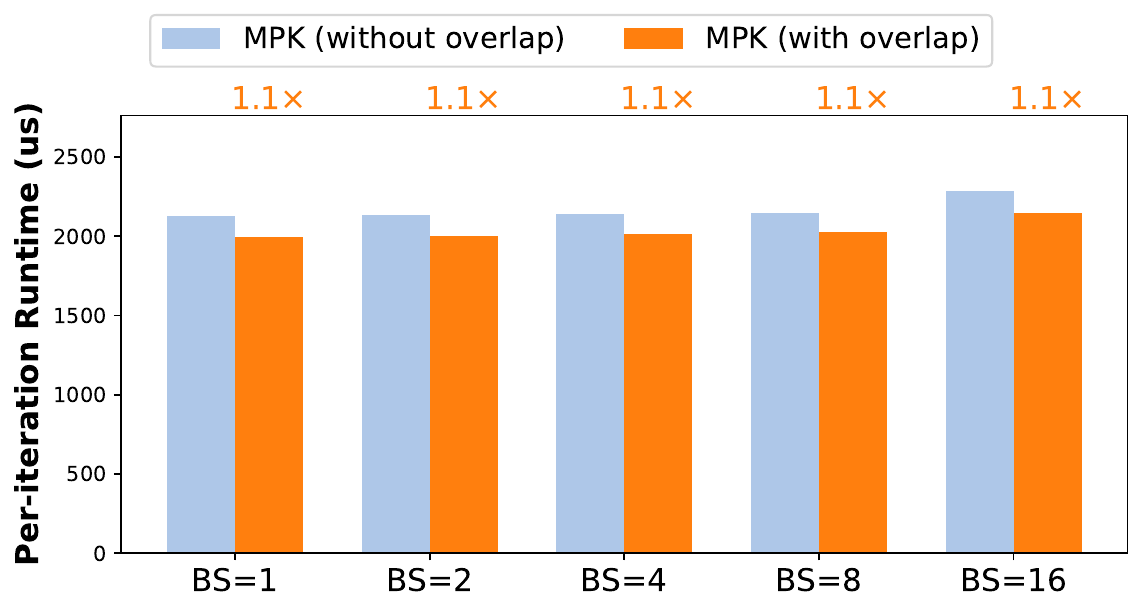}
    \vspace{\captionvspace}
    \caption{Ablation study on compute--communication overlap. \rev{We measure the runtime of Qwen3-1.7B on four H100 GPUs using tensor parallelism; lower is better. We compare \sys with compute--communication overlap enabled and disabled to quantify the benefit of fine-grained overlap.}}
    \vspace{\captionvspace}
    \label{fig:ablation_multigpu}
\end{figure}

\paragraph{Compute-communication overlap.} \sys captures fine-grained dependencies between tasks (\Cref{fig:compiler_workflow}), allowing the runtime to opportunistically overlap compute and communication. \Cref{fig:ablation_multigpu} evaluates the impact of this optimization. To disable overlap, we restrict the \graph to capture only coarse-grained dependencies between each collective operator, such as {\tt AllReduce}, and its preceding computation using a single event, as illustrated in \Cref{fig:compiler_workflow}c. Enabling fine-grained compute--communication overlap reduces per-iteration latency by 1.1$\times$.

\paragraph{Kernel-launch reduction.} \Sys executes the entire model with a single kernel launch, whereas a kernel-per-operator execution of Qwen3-8B issues 293 kernel launches per token. On B200, each launch costs 3.8~$\mu$s in eager execution, totaling 1.1~ms per token, and 0.8~$\mu$s with CUDA Graphs, totaling 0.2~ms per token. \Sys avoids this overhead; its in-kernel scheduler accounts for only 0.28\% of total runtime (\Cref{sec:impl}).

\begin{revblock}
\subsection{Compiler-Stage Ablation}
\label{subsec:compiler_ablation}
The preceding ablations isolate two runtime mechanisms. We now isolate the contribution of each compiler stage described in \Cref{sec:compiler}. Because these stages serve different purposes, we report the most appropriate metric for each stage rather than using a single common metric. \Cref{tab:compiler_ablation} summarizes the results on three representative models.

\paragraph{Operator decomposition.}
Operator decomposition partitions each operator into independent SM-level tasks. As shown in \Cref{tab:compiler_ablation}, decomposition exposes substantial parallelism in real models: each operator is split into 32--47 tasks on average. For example, in Qwen3-8B, 293 operators are decomposed into 13{,}867 tasks. Compute-intensive operators, including linear layers, attention, and MoE experts, each expose tens of parallel tasks, while only a small number of pointwise operators, such as normalization at batch size one, map to a single task. This decomposition provides the runtime with sufficient independent work to keep all SMs busy.

\paragraph{Event fusion.} Without fusion, dependency analysis emits a separate event for every overlapping producer--consumer task pair. Successor-set and predecessor-set fusion collapse events that share the same consumer set or producer set into a single synchronization point. This optimization is highly effective: across the evaluated models, the final \graphs contain only 1{,}142--2{,}366 events, yet these events encode 69{,}000--162{,}000 producer--consumer task-pair dependencies. This corresponds to a 37--118$\times$ reduction in synchronization events (\Cref{tab:compiler_ablation}). The reduction is largest for the MoE model because expert routing creates many-to-many dependencies that are especially amenable to fusion.

\paragraph{\graph normalization.}
\graph normalization inserts auxiliary tasks and events only when a task triggers or depends on more than one event, namely at forks and joins. The compiled LLM forward \graphs in our evaluation are almost entirely sequential: we observe no fork/join groups in all three compiled \graphs because operators that would otherwise fan out, such as the query/key/value projections, are emitted as fused operators. Normalization therefore leaves these \graphs essentially unchanged, confirming the ``deep, not wide'' structure discussed in \Cref{sec:compiler}. Nevertheless, normalization remains necessary for correctness when parallel branches do occur, such as in the unfused Q/K/V example in \Cref{fig:compiler_workflow}. In those cases, normalization bounds each task's event fan-in and fan-out to one, preserving the fixed-size, indirection-free encoding.

\paragraph{\graph linearization.}
\graph linearization encodes each event's successor tasks as a contiguous index range rather than an explicit task list. We measure the resulting on-device footprint as the size of the successor encoding with and without linearization. The contiguous-range encoding reduces this footprint by 4.4--5.9$\times$ on dense models, such as from 110{,}932 bytes to 18{,}928 bytes for Qwen3-8B, and by 15.0$\times$ on the MoE model. The MoE model benefits most because high-fan-out expert-routing events are particularly compact under the range-based encoding. Linearization achieves these reductions without losing dependency information.

\begin{table}
    \centering
    \caption{\rev{Per-compiler-stage statistics on B200. {\bf Ops}: number of operators; {\bf Tasks/op}: average number of tasks per operator after decomposition; {\bf Events}: number of events in the final \graph; {\bf Fusion}: event-count reduction from event fusion; {\bf Lin.}: device-memory footprint reduction from linearization.}}

    \small 
    \label{tab:compiler_ablation}
    \setlength{\tabcolsep}{4pt}
    \begin{tabular}{l|rrrrr}
    \toprule
    {\bf Model} & {\bf Ops} & {\bf Tasks/op} & {\bf Events} & {\bf Fusion} & {\bf Lin.}\\
    \midrule
    Qwen3-1.7B    & 229 & 35.6 & 1{,}870 & 37$\times$  & 4.4$\times$ \\
    Qwen3-8B      & 293 & 47.3 & 2{,}366 & 68$\times$  & 5.9$\times$ \\
    Qwen3-30B-A3B & 533 & 32.2 & 1{,}142 & 118$\times$ & 15.0$\times$ \\
    \bottomrule
    \end{tabular}
    \vspace{\captionvspace}
\end{table}

\end{revblock}

\begin{revblock}
\section{Discussion}
\label{sec:discussion}

\paragraph{Resource footprint.}
A common concern with mega-kernels is that a single kernel must reserve, on every SM, the combined resources required by all fused operators. 
\Sys avoids this issue: tasks are time-multiplexed across SMs rather than statically partitioned, so each SM uses only the resources required by the task it is currently executing (\Cref{sec:runtime}). Shared memory follows the same pattern: the paged abstraction (\Cref{subsec:runtime_optimizations}) acquires and releases pages at task granularity. The main exception is per-thread register usage, which is fixed for the mega-kernel at the maximum required across task types.

\paragraph{Integrating hand-tuned kernels and porting to new hardware.}
Each \sys task is a CUDA device function with a uniform calling convention defined by the \graph. This interface gives \sys two forms of flexibility. First, a hand-tuned kernel can be incorporated into \sys by wrapping it as the device function for a task; the task then executes under the same event-driven schedule as compiler-generated tasks, without requiring changes to the runtime
or the \graph. \Sys can thus reuse hand-optimized implementations when  available while automatically generating the remaining tasks. Second, porting \sys to a new GPU architecture only requires updating the per-task code generators that emit these device functions; the \graph representation, the compiler's graph-level transformations, the runtime, and the scheduler remain unchanged. Therefore, per-task code generation is the only hardware-specific component of \sys (\Cref{sec:compiler}), while the rest of the system is shared
across GPU architectures.
\end{revblock}
\section{Related Work}
\label{sec:related}

\paragraph{Manually designed kernels.}
Existing ML frameworks such as TensorFlow XLA~\cite{tensorflow_xla, Tensorflow}, PyTorch~\cite{pytorch}, and TensorRT~\cite{tensorrt} adopt a kernel-per-operator approach and rely on GPU experts to manually design and implement kernels for individual operators.
For attention alone, various specialized kernels have been developed, including FlashAttention~\cite{dao2023flash, tri2023flashdecoding, hong2024flashdecoding}, FasterTransformer~\cite{fastertransformer}, and FlashInfer~\cite{ye2025flashinfer}, each targeting specific architectural features or deployment scenarios. Current systems rely on many specialized kernel libraries, making it hard to unify the entire inference pipeline into one mega-kernel.

\paragraph{ML compilers.}
A large body of work has explored {\em compiler-based} generation of high-performance kernels for tensor programs. Systems such as TVM~\cite{tvm, tvm_auto_tuner}, Ansor~\cite{ansor}, and Triton~\cite{tillet2019triton}, alongside others~\cite{zheng2020flextensor, hagedorn2023graphene, feng2022tensorir, hu2024korch}, build on the algorithm–schedule separation introduced by Halide~\cite{halide, autohalide}.
Another line of work employs {\em superoptimization} techniques to automatically search for efficient kernel implementations from high-level specifications~\cite{TASO, wang2021pet, zheng2023einnet, tensat, unger2022unity}.
However, these compilers are largely designed around operator- or graph-level optimization and do not support generating a unified mega-kernel or coordinating cross-operator execution.

\paragraph{Mega-kernels.}
Prior efforts on mega-kernels largely rely on manual design. For example, FlashDMoE fuses mixture-of-experts computation with inter-GPU communication into a single handcrafted mega-kernel~\cite{aimuyo2025flashdmoe}. Spector et al.\ manually developed a low-latency mega-kernel for LLaMA-1B~\cite{touvron2023llama, megakernel}.
These approaches require extensive engineering effort and deep GPU expertise, and they do not generalize across models or GPUs.
In contrast, \sys provides a compiler-based solution that automatically transforms a tensor program into a highly optimized mega-kernel, eliminating the need for manual fusion or hand-written mega-kernel implementations.

\rev{A related line of work simplifies the development of manually fused kernels. ThunderKittens~\cite{spector2024thunderkittens} provides tile-based abstractions for high-performance kernels, and PipeThreader~\cite{pipethreader} exposes software-pipelined kernel execution to the programmer. These tools still require the developer to decide which operators to fuse, how to pipeline them, and how to coordinate their execution. \Sys instead compiles an entire tensor program into SM-level tasks and executes them as one mega-kernel. These abstractions are therefore complementary to \sys: they could be used to implement individual task kernels within the \sys runtime. }

\rev{TileRT~\cite{tilert} similarly decomposes operators into fine-grained tile-level tasks executed as a single persistent kernel, validating our core idea that persistent mega-kernel execution is key to low-latency inference.}

\begin{revblock}\paragraph{Compute--communication overlap.}
A large body of work overlaps collective communication with computation to hide
inter-GPU communication latency. DeepEP~\cite{deepep} provides expert-parallel
all-to-all communication kernels for MoE models, and
Triton-Distributed~\cite{tritondistributed} extends the Triton compiler to express
overlapping distributed kernels. TokenWeave~\cite{tokenweave},
FlashOverlap~\cite{flashoverlap}, and ParallelKittens~\cite{parallelkitten} fuse or
co-schedule communication with dependent computation to reduce exposed
communication latency. Earlier work from Google decomposes collective operators so that they overlap with dependent computation~\cite{wang2023overlap}. These systems realize overlap at the granularity of
individual kernels or collective operators. In contrast, \sys represents communication and computation uniformly as tasks in one \graph and schedules them using the same in-kernel runtime (\Cref{subsec:multi_gpu}). Compute--communication overlap therefore emerges from the global task schedule and applies uniformly to both tensor-parallel collectives and expert-parallel all-to-all communication.\end{revblock}

\section{Conclusion}
\label{sec:conclusion}


This paper presents \sys, the first compiler and runtime system that automatically transforms multi-GPU model inference into a fully fused mega-kernel. By introducing SM-level task graphs and an in-kernel parallel runtime, \sys overcomes key limitations of the kernel-per-operator execution model, enabling inter-operator software pipelining, fine-grained overlap of computation and communication, and the elimination of kernel-launch and CPU-side scheduling overheads. Our evaluation shows that \sys brings LLM serving latency close to hardware limits and significantly improves throughput across models and GPU generations. By unifying execution within a single mega-kernel while preserving compatibility with existing ML frameworks, \sys opens a new path toward high-performance, compiler-driven inference systems.

\section*{Acknowledgment}
We thank the anonymous OSDI reviewers and our shepherd for their valuable comments and suggestions. This work was partially supported by NSF awards CNS-2211882 and CNS-2239351, a Sloan Research Fellowship, and research awards from Amazon, Cisco, Google, Jane Street, Meta, NVIDIA, Oracle, Qualcomm, and Samsung. We also gratefully acknowledge NVIDIA for providing access to a DGX B200 system.

\balance
\bibliography{bibliography}

@misc{nccl,
Howpublished={\url{https://developer.nvidia.com/nccl}},
Title = {NVIDIA NCCL},
year={2021}
}

@misc{megakernel,
Howpublished={\url{https://hazyresearch.stanford.edu/blog/2025-05-27-no-bubbles}},
authors = {Benjamin Spector and Jordan Juravsky and Stuart Sul and Owen Dugan and Dylan Lim and Dan Fu and Simran Arora and Chris Ré},
Title = {{Designing a Low-Latency Megakernel for Llama-1B}},
year={2025}
}

@misc{moegroupedgemm,
Howpublished={\href{https://pytorch.org/blog/accelerating-moes-with-a-triton-persistent-cache-aware-grouped-gemm-kernel/}{Link}},
authors = {Less Wright and Adnan Hoque and Garrett Goon},
Title = {{Accelerating MoE’s with a Triton Persistent Cache-Aware Grouped GEMM Kernel}},
year={2025}
}

@software{jax2018github,
  author = {James Bradbury and Roy Frostig and Peter Hawkins and Matthew James Johnson and Chris Leary and Dougal Maclaurin and George Necula and Adam Paszke and Jake Vander{P}las and Skye Wanderman-{M}ilne and Qiao Zhang},
  title = {{JAX}: composable transformations of {P}ython+{N}um{P}y programs},
  url = {http://github.com/jax-ml/jax},
  version = {0.3.13},
  year = {2018},
}

@misc{aimuyo2025flashdmoe,
      title={FlashDMoE: Fast Distributed MoE in a Single Kernel}, 
      author={Osayamen Jonathan Aimuyo and Byungsoo Oh and Rachee Singh},
      year={2025},
      eprint={2506.04667},
      archivePrefix={arXiv},
      primaryClass={cs.DC},
      url={https://arxiv.org/abs/2506.04667}, 
}

@misc{nvidia-pdl,
Howpublished={\url{https://docs.nvidia.com/cuda/cuda-c-programming-guide/index.html}},
Title = {{Programmatic Dependent Launch}},
year={2025}
}

@misc{nvshmem,
Howpublished={\url{https://docs.nvidia.com/nvshmem/api/index.html}},
Title = {{NVIDIA OpenSHMEM Library (NVSHMEM) Documentation}},
year={2025}
}

@misc{huggingfacemodels,
Howpublished={\url{https://huggingface.co/models}},
Title = {{Huggingface Models}},
year={2023}
}

@inproceedings{kwon2023vllm,
  title={Efficient Memory Management for Large Language Models},
  author={Kwon, Woosuk and Li, Zhuohan and Liu, Siyuan and Wu, Xin and Zeng, Michael and Zhang, Xiangru and Zou, Yuhao and Moritz, Scott},
  booktitle={Proceedings of the ACM Symposium on Operating Systems Principles (SOSP)},
  year={2023}
}

@inproceedings{zheng2024sglang,
author = {Zheng, Lianmin and Yin, Liangsheng and Xie, Zhiqiang and Sun, Chuyue and Huang, Jeff and Yu, Cody Hao and Cao, Shiyi and Kozyrakis, Christos and Stoica, Ion and Gonzalez, Joseph E. and Barrett, Clark and Sheng, Ying},
title = {SGLang: efficient execution of structured language model programs},
year = {2024},
isbn = {9798331314385},
publisher = {Curran Associates Inc.},
address = {Red Hook, NY, USA},
booktitle = {Proceedings of the 38th International Conference on Neural Information Processing Systems},
articleno = {2000},
numpages = {27},
location = {Vancouver, BC, Canada},
series = {NIPS '24}
}

@inproceedings{zhu2025nanoflow,
  title={$\{$NanoFlow$\}$: Towards optimal large language model serving throughput},
  author={Zhu, Kan and Gao, Yufei and Zhao, Yilong and Zhao, Liangyu and Zuo, Gefei and Gu, Yile and Xie, Dedong and Ye, Zihao and Kamahori, Keisuke and Lin, Chien-Yu and others},
  booktitle={19th USENIX Symposium on Operating Systems Design and Implementation (OSDI 25)},
  pages={749--765},
  year={2025}
}

@inproceedings{unger2022unity,
  author    = {Colin Unger and
               Zhihao Jia and
               Wei Wu and
               Sina Lin and
               Mandeep Baines and
               Carlos Efrain Quintero Narvaez and
               Vinay Ramakrishnaiah and
               Nirmal Prajapati and
               Patrick S. McCormick and
               Jamaludin Mohd{-}Yusof and
               Xi Luo and
               Dheevatsa Mudigere and
               Jongsoo Park and
               Misha Smelyanskiy and
               Alex Aiken},
  title     = {Unity: Accelerating {DNN} Training Through Joint Optimization of Algebraic
               Transformations and Parallelization},
  booktitle = {16th {USENIX} Symposium on Operating Systems Design and Implementation,
               {OSDI} 2022, Carlsbad, CA, USA, July 11-13, 2022},
  pages     = {267--284},
  publisher = {{USENIX} Association},
  year      = {2022},
  url       = {https://www.usenix.org/conference/osdi22/presentation/unger},
  bibsource = {dblp computer science bibliography, https://dblp.org}
}

@misc{ye2025flashinfer,
      title={FlashInfer: Efficient and Customizable Attention Engine for LLM Inference Serving}, 
      author={Zihao Ye and Lequn Chen and Ruihang Lai and Wuwei Lin and Yineng Zhang and Stephanie Wang and Tianqi Chen and Baris Kasikci and Vinod Grover and Arvind Krishnamurthy and Luis Ceze},
      year={2025},
      eprint={2501.01005},
      archivePrefix={arXiv},
      primaryClass={cs.DC},
      url={https://arxiv.org/abs/2501.01005}, 
}

@inproceedings{wu2025mirage,
  title={Mirage: A $\{$Multi-Level$\}$ Superoptimizer for Tensor Programs},
  author={Wu, Mengdi and Cheng, Xinhao and Liu, Shengyu and Shi, Chunan and Ji, Jianan and Ao, Man Kit and Velliengiri, Praveen and Miao, Xupeng and Padon, Oded and Jia, Zhihao},
  booktitle={19th USENIX Symposium on Operating Systems Design and Implementation (OSDI 25)},
  pages={21--38},
  year={2025}
}

@inproceedings {wang2021pet,
author = {Haojie Wang and Jidong Zhai and Mingyu Gao and Zixuan Ma and Shizhi Tang and Liyan Zheng and Yuanzhi Li and Kaiyuan Rong and Yuanyong Chen and Zhihao Jia},
title = {{PET}: Optimizing Tensor Programs with Partially Equivalent Transformations and Automated Corrections},
booktitle = {15th USENIX Symposium on Operating Systems Design and Implementation (OSDI 21)},
year = {2021},
isbn = {978-1-939133-22-9},
pages = {37--54},
url = {https://www.usenix.org/conference/osdi21/presentation/wang},
publisher = {USENIX Association},
month = jul,
}

@article{ansor,
  author    = {Lianmin Zheng and
               Chengfan Jia and
               Minmin Sun and
               Zhao Wu and
               Cody Hao Yu and
               Ameer Haj{-}Ali and
               Yida Wang and
               Jun Yang and
               Danyang Zhuo and
               Koushik Sen and
               Joseph E. Gonzalez and
               Ion Stoica},
  title     = {Ansor : Generating High-Performance Tensor Programs for Deep Learning},
  journal   = {CoRR},
  volume    = {abs/2006.06762},
  year      = {2020},
  url       = {https://arxiv.org/abs/2006.06762},
  archivePrefix = {arXiv},
  eprint    = {2006.06762},
  bibsource = {dblp computer science bibliography, https://dblp.org}
}

@misc{tri2023flashdecoding,
Howpublished={\url{https://crfm.stanford.edu/2023/10/12/flashdecoding.html}},
Title = {Flash-Decoding for long-context inference},
year={2023}
}

@misc{fastertransformer,
Howpublished={\url{https://github.com/NVIDIA/FasterTransformer}},
Title = {Transformer related optimizations},
year={2020}
}

@misc{triton-flashattention,
Howpublished={\url{https://triton-lang.org/main/getting-started/tutorials/06-fused-attention.html}},
Title = {A {Triton} implementation of the {FlashAttention2} algorithm},
year={2023}
}

@misc{cutlass,
Howpublished={\url{https://github.com/NVIDIA/cutlass}},
Title = {NVIDIA/cutlass: CUDA Templates for Linear Algebra Subroutines},
year={2019}
}

@inproceedings{TASO,
author = {Jia, Zhihao and Padon, Oded and Thomas, James and Warszawski, Todd and Zaharia, Matei and Aiken, Alex},
title = {TASO: Optimizing Deep Learning Computation with Automatic Generation of Graph Substitutions},
year = {2019},
isbn = {9781450368735},
publisher = {Association for Computing Machinery},
address = {New York, NY, USA},
url = {https://doi.org/10.1145/3341301.3359630},
doi = {10.1145/3341301.3359630},
booktitle = {Proceedings of the 27th ACM Symposium on Operating Systems Principles},
pages = {47–62},
numpages = {16},
location = {Huntsville, Ontario, Canada},
series = {SOSP '19}
}

@article{Megatron,
  author    = {Mohammad Shoeybi and
               Mostofa Patwary and
               Raul Puri and
               Patrick LeGresley and
               Jared Casper and
               Bryan Catanzaro},
  title     = {Megatron-LM: Training Multi-Billion Parameter Language Models Using
               Model Parallelism},
  journal   = {CoRR},
  volume    = {abs/1909.08053},
  year      = {2019},
  url       = {http://arxiv.org/abs/1909.08053},
  archivePrefix = {arXiv},
  eprint    = {1909.08053},
  bibsource = {dblp computer science bibliography, https://dblp.org}
}

@incollection{tvm_auto_tuner,
title = {Learning to Optimize Tensor Programs},
author = {Chen, Tianqi and Zheng, Lianmin and Yan, Eddie and Jiang, Ziheng and Moreau, Thierry and Ceze, Luis and Guestrin, Carlos and Krishnamurthy, Arvind},
booktitle = {Advances in Neural Information Processing Systems 31},
series = {NeurIPS'18},
year = {2018},
}

@misc{hong2024flashdecoding,
      title={FlashDecoding++: Faster Large Language Model Inference on GPUs}, 
      author={Ke Hong and Guohao Dai and Jiaming Xu and Qiuli Mao and Xiuhong Li and Jun Liu and Kangdi Chen and Yuhan Dong and Yu Wang},
      year={2024},
      eprint={2311.01282},
      archivePrefix={arXiv},
      primaryClass={cs.LG}
}

@inproceedings{tillet2019triton,
author = {Tillet, Philippe and Kung, H. T. and Cox, David},
title = {Triton: an intermediate language and compiler for tiled neural network computations},
year = {2019},
isbn = {9781450367196},
publisher = {Association for Computing Machinery},
address = {New York, NY, USA},
url = {https://doi.org/10.1145/3315508.3329973},
doi = {10.1145/3315508.3329973},
booktitle = {Proceedings of the 3rd ACM SIGPLAN International Workshop on Machine Learning and Programming Languages},
pages = {10–19},
numpages = {10},
location = {Phoenix, AZ, USA},
series = {MAPL 2019}
}

@article{autohalide,
 author = {Mullapudi, Ravi Teja and Adams, Andrew and Sharlet, Dillon and Ragan-Kelley, Jonathan and Fatahalian, Kayvon},
 title = {Automatically Scheduling Halide Image Processing Pipelines},
 journal = {ACM Trans. Graph.},
 issue_date = {July 2016},
 volume = {35},
 number = {4},
 year = {2016},
}

@inproceedings{halide,
 author = {Ragan-Kelley, Jonathan and Barnes, Connelly and Adams, Andrew and Paris, Sylvain and Durand, Fr{\'e}do and Amarasinghe, Saman},
 title = {Halide: A Language and Compiler for Optimizing Parallelism, Locality, and Recomputation in Image Processing Pipelines},
 booktitle = {Proceedings of the 34th ACM SIGPLAN Conference on Programming Language Design and Implementation},
 series = {PLDI '13},
 year = {2013},
}

@misc{tensorrt,
key = {TensorRT},
Howpublished={\url{https://developer.nvidia.com/tensorrt}},
Title = {{NVIDIA TensorRT}: Programmable Inference Accelerator},
year = {2017}
}

@misc{tensorflow_xla,
Howpublished={\url{https://www.tensorflow.org/xla}},
Title = {XLA: Optimizing Compiler for TensorFlow},
year={2017}
}

@misc{feng2022tensorir,
      title={TensorIR: An Abstraction for Automatic Tensorized Program Optimization}, 
      author={Siyuan Feng and Bohan Hou and Hongyi Jin and Wuwei Lin and Junru Shao and Ruihang Lai and Zihao Ye and Lianmin Zheng and Cody Hao Yu and Yong Yu and Tianqi Chen},
      year={2022},
      eprint={2207.04296},
      archivePrefix={arXiv},
      primaryClass={cs.LG}
}

@inproceedings{hagedorn2023graphene,
author = {Hagedorn, Bastian and Fan, Bin and Chen, Hanfeng and Cecka, Cris and Garland, Michael and Grover, Vinod},
title = {Graphene: An IR for Optimized Tensor Computations on GPUs},
year = {2023},
isbn = {9781450399180},
publisher = {Association for Computing Machinery},
address = {New York, NY, USA},
url = {https://doi.org/10.1145/3582016.3582018},
doi = {10.1145/3582016.3582018},
booktitle = {Proceedings of the 28th ACM International Conference on Architectural Support for Programming Languages and Operating Systems, Volume 3},
pages = {302–313},
numpages = {12},
location = {, Vancouver, BC, Canada, },
series = {ASPLOS 2023}
}

@inproceedings {zheng2023einnet,
author = {Liyan Zheng and Haojie Wang and Jidong Zhai and Muyan Hu and Zixuan Ma and Tuowei Wang and Shuhong Huang and Xupeng Miao and Shizhi Tang and Kezhao Huang and Zhihao Jia},
title = {{EINNET}: Optimizing Tensor Programs with {Derivation-Based} Transformations},
booktitle = {17th USENIX Symposium on Operating Systems Design and Implementation (OSDI 23)},
year = {2023},
isbn = {978-1-939133-34-2},
address = {Boston, MA},
pages = {739--755},
url = {https://www.usenix.org/conference/osdi23/presentation/zheng},
publisher = {USENIX Association},
month = jul
}

@inproceedings{zheng2020flextensor,
author = {Zheng, Size and Liang, Yun and Wang, Shuo and Chen, Renze and Sheng, Kaiwen},
title = {FlexTensor: An Automatic Schedule Exploration and Optimization Framework for Tensor Computation on Heterogeneous System},
year = {2020},
isbn = {9781450371025},
publisher = {Association for Computing Machinery},
address = {New York, NY, USA},
url = {https://doi.org/10.1145/3373376.3378508},
doi = {10.1145/3373376.3378508},
booktitle = {Proceedings of the Twenty-Fifth International Conference on Architectural Support for Programming Languages and Operating Systems},
pages = {859–873},
numpages = {15},
location = {Lausanne, Switzerland},
series = {ASPLOS '20}
}

@article{tvm,
  author    = {Tianqi Chen and
               Thierry Moreau and
               Ziheng Jiang and
               Haichen Shen and
               Eddie Q. Yan and
               Leyuan Wang and
               Yuwei Hu and
               Luis Ceze and
               Carlos Guestrin and
               Arvind Krishnamurthy},
  title     = {{TVM:} End-to-End Optimization Stack for Deep Learning},
  journal   = {CoRR},
  volume    = {abs/1802.04799},
  year      = {2018},
  url       = {http://arxiv.org/abs/1802.04799},
}

@inproceedings{Tensorflow,
  title={TensorFlow: A System for Large-Scale Machine Learning.},
  author={Abadi, Mart\'{\i}n and Barham, Paul and Chen, Jianmin and Chen, Zhifeng and Davis, Andy and Dean, Jeffrey and Devin, Matthieu and Ghemawat, Sanjay and Irving, Geoffrey and Isard, Michael and Kudlur, Manjunath and Levenberg, Josh and Monga, Rajat and Moore, Sherry and Murray, Derek G. and Steiner, Benoit and Tucker, Paul and Vasudevan, Vijay and Warden, Pete and Wicke, Martin and Yu, Yuan and Zheng, Xiaoqiang},
  booktitle = {Proceedings of the 12th USENIX Conference on Operating Systems Design and Implementation},
  series={OSDI},
  year={2016}
}

@misc{pytorch,
Key = {PyTorch},
Howpublished = {\url{https://pytorch.org}},
Title = {{Tensors and Dynamic neural networks in Python with strong GPU acceleration.}},
year={2017}
}

@article{tensat,
  title = {Equality {{Saturation}} for {{Tensor Graph Superoptimization}}},
  author = {Yang, Yichen and Phothilimthana, Phitchaya and Wang, Yisu and Willsey, Max and Roy, Sudip and Pienaar, Jacques},
  year = {2021},
  month = mar,
  journal = {Proceedings of Machine Learning and Systems},
  volume = {3},
  pages = {255--268},
  langid = {english}
}

@article{touvron2023llama,
  title={Llama: Open and efficient foundation language models},
  author={Touvron, Hugo and Lavril, Thibaut and Izacard, Gautier and Martinet, Xavier and Lachaux, Marie-Anne and Lacroix, Timoth{\'e}e and Rozi{\`e}re, Baptiste and Goyal, Naman and Hambro, Eric and Azhar, Faisal and others},
  journal={arXiv preprint arXiv:2302.13971},
  year={2023}
}

@inproceedings{yu2022orca,
  title={Orca: A distributed serving system for $\{$Transformer-Based$\}$ generative models},
  author={Yu, Gyeong-In and Jeong, Joo Seong and Kim, Geon-Woo and Kim, Soojeong and Chun, Byung-Gon},
  booktitle={16th USENIX Symposium on Operating Systems Design and Implementation (OSDI 22)},
  pages={521--538},
  year={2022}
}

@Misc{transformers,
    title = {Transformers: State-of-the-art Machine Learning for Pytorch, TensorFlow, and JAX.},
    author = {Thomas Wolf and Lysandre Debut and Victor Sanh and Julien Chaumond and Clement Delangue and Anthony Moi and Pierric Cistac and Tim Rault and Rémi Louf and Morgan Funtowicz and Joe Davison and Sam Shleifer and Patrick von Platen and Clara Ma and Yacine Jernite and Julien Plu and Canwen Xu and Teven Le Scao and Sylvain Gugger and Mariama Drame and Quentin Lhoest and Alexander M. Rush},
    howpublished = {\url{https://github.com/huggingface/transformers}},
  year =         {2022}
}

@misc{dao2023flash,
  title={Flash-Decoding for Long-Context Inference},
  author={Dao, Tri and Haziza, Daniel and Massa, Francisco and Sizov, Grigory},
  year={2023}
}

@inproceedings{hu2024korch,
author = {Hu, Muyan and Venkatram, Ashwin and Biswas, Shreyashri and Marimuthu, Balamurugan and Hou, Bohan and Oliaro, Gabriele and Wang, Haojie and Zheng, Liyan and Miao, Xupeng and Zhai, Jidong and Jia, Zhihao},
title = {Optimal Kernel Orchestration for Tensor Programs with Korch},
year = {2024},
isbn = {9798400703867},
publisher = {Association for Computing Machinery},
address = {New York, NY, USA},
url = {https://doi.org/10.1145/3620666.3651383},
doi = {10.1145/3620666.3651383},
booktitle = {Proceedings of the 29th ACM International Conference on Architectural Support for Programming Languages and Operating Systems, Volume 3},
pages = {755–769},
numpages = {15},
location = {La Jolla, CA, USA},
series = {ASPLOS '24}
}

@inproceedings{spector2024thunderkittens,
  title     = {ThunderKittens: Simple, Fast, and Adorable AI Kernels},
  author    = {Spector, Benjamin F. and Arora, Simran and Singhal, Aaryan and Parthasarathy, Arjun and Fu, Daniel Y. and R\'e, Christopher},
  booktitle = {International Conference on Learning Representations (ICLR)},
  year      = {2025}
}

@inproceedings{pipethreader,
  title     = {PipeThreader: Software-Defined Pipelining for Efficient DNN Execution},
  author    = {Cheng, Yu and Wang, Lei and Shi, Yining and Xia, Yuqing and Ma, Lingxiao and Xue, Jilong and Wang, Yang and Mo, Zhiwen and Chen, Feiyang and Yang, Fan and Yang, Mao and Yang, Zhi},
  booktitle = {19th USENIX Symposium on Operating Systems Design and Implementation (OSDI)},
  year      = {2025}
}

@misc{deepep,
  title        = {{DeepEP}: An Efficient Expert-Parallel Communication Library},
  author       = {{DeepSeek-AI}},
  year         = {2025},
  howpublished = {\url{https://github.com/deepseek-ai/DeepEP}}
}

@misc{tritondistributed,
  title         = {Triton-distributed: Programming Overlapping Kernels on Distributed AI Systems with the Triton Compiler},
  author        = {Zheng, Size and Bao, Wenlei and Hou, Qi and Zheng, Xuegui and Fang, Jin and Huang, Chenhui and Li, Tianqi and Duanmu, Haojie and Chen, Renze and Xu, Ruifan and Guo, Yifan and Zheng, Ningxin and Jiang, Ziheng and Di, Xinyi and Wang, Dongyang and Ye, Jianxi and Lin, Haibin and Chang, Li-Wen and Lu, Liqiang and Liang, Yun and Zhai, Jidong and Liu, Xin},
  year          = {2025},
  eprint        = {2504.19442},
  archivePrefix = {arXiv},
  primaryClass  = {cs.DC},
  howpublished  = {arXiv preprint arXiv:2504.19442}
}

@inproceedings{tokenweave,
  title     = {{TokenWeave}: Efficient Compute-Communication Overlap for Distributed LLM Inference},
  author    = {Gond, Raja and Kwatra, Nipun and Ramjee, Ramachandran},
  booktitle = {Proceedings of Machine Learning and Systems (MLSys)},
  year      = {2026}
}

@inproceedings{parallelkitten,
  title     = {{ParallelKittens}: Systematic and Practical Simplification of Multi-GPU AI Kernels},
  author    = {Sul, Stuart H. and Arora, Simran and Spector, Benjamin F. and R\'e, Christopher},
  booktitle = {Proceedings of Machine Learning and Systems (MLSys)},
  year      = {2026}
}

@inproceedings{flashoverlap,
  title     = {{FlashOverlap}: A Lightweight Design for Efficiently Overlapping Communication and Computation},
  author    = {Hong, Ke and Li, Xiuhong and Liu, Minxu and Mao, Qiuli and Wu, Tianqi and Huang, Zixiao and Chen, Lufang and Wang, Zhong and Zhang, Yichong and Zhu, Zhenhua and Dai, Guohao and Wang, Yu},
  booktitle = {Proceedings of the European Conference on Computer Systems (EuroSys)},
  year      = {2026}
}

@inproceedings{wang2023overlap,
  title     = {Overlap Communication with Dependent Computation via Decomposition in Large Deep Learning Models},
  author    = {Wang, Shibo and Wei, Jinliang and Sabne, Amit and Davis, Andy and Ilbeyi, Berkin and Hechtman, Blake and Chen, Dehao and Murthy, Karthik Srinivasa and Maggioni, Marcello and Zhang, Qiao and Kumar, Sameer and Guo, Tongfei and Xu, Yuanzhong and Zhou, Zongwei},
  booktitle = {Proceedings of the 28th ACM International Conference on Architectural Support for Programming Languages and Operating Systems, Volume 1 (ASPLOS)},
  pages     = {93--106},
  year      = {2023},
  doi       = {10.1145/3567955.3567959}
}

@misc{tilert,
    title        = {{TileRT}: Tile-Based Runtime for Ultra-Low-Latency LLM Inference},
    author       = {{Tile-AI}},
    year         = {2026},
    howpublished = {\url{https://github.com/tile-ai/TileRT}}
  }
\bibliographystyle{plain}

\newpage
\appendix
\begin{revblock}
\section{Artifact Appendix}

\subsection*{Abstract}
The artifact is the \sys compiler and runtime together with scripts that reproduce
the paper's end-to-end latency results. \sys transforms a PyTorch model into a
single mega-kernel; the scripts measure per-token decoding latency for \sys,
PyTorch, vLLM, and SGLang across five LLMs, five batch sizes, and three GPU
generations.

\subsection*{Scope}
The artifact validates the paper's main performance claims: (i) on single-batch
serving, \sys reduces per-token latency by 1.0--1.7$\times$ over vLLM and SGLang,
with the largest gains on smaller models and newer GPUs (\Cref{fig:end_to_end});
(ii) under tensor parallelism, \sys improves throughput by up to 10$\times$ over
PyTorch and 1.1--1.4$\times$ over vLLM/SGLang on 8 H100s
(\Cref{fig:end_to_end_multigpu}); and (iii) cross-task pipelining and
compute--communication overlap each contribute measurable speedups
(\Cref{fig:pipe_ablation,fig:ablation_multigpu}).

\subsection*{Contents}
The repository holds the \sys compiler and in-kernel runtime, a runnable demo for
each model under \texttt{demo/}, and one evaluation driver per GPU under
\texttt{artifact\_evaluation/}. A driver sweeps the five models over batch sizes
1 to 16 for all four systems and writes one result file per run, recording the
system, GPU, model, batch size, and measured per-token latency. Collecting these
files reproduces \Cref{fig:end_to_end} and the ablation figures. A
\texttt{setup.sh} script installs the dependencies and builds \sys.

\subsection*{Hosting}
The artifact is on the frozen branch \texttt{tgx-osdi26-ae} of
\url{https://github.com/mirage-project/mirage} (commit \texttt{8b981a4}),
archived on Zenodo at \url{https://doi.org/10.5281/zenodo.20563064}
(DOI:~10.5281/zenodo.20563064).

\subsection*{Requirements}
\noindent\textbf{Hardware.} One NVIDIA A100, H100 (SXM), or B200 for the single-GPU
sweeps, and a multi-GPU node (4--8 H100s) for the tensor-parallel experiments. The
four smaller models fit on a 40\,GB A100; Qwen3-30B-A3B needs at least 80\,GB and
runs on an 80\,GB A100, an H100, or a B200.

\noindent\textbf{Software.} Linux with CUDA (12.8 in the paper), PyTorch~2.7,
NVSHMEM and OpenMPI for the multi-GPU runs, and \texttt{transformers}~4.57.1.
\texttt{setup.sh} installs these and builds \sys; vLLM and SGLang each run in a
separate virtual environment.

\noindent\textbf{Benchmarks.} Qwen3-0.6B, Llama-3.2-1B-Instruct, Qwen3-1.7B,
Qwen3-8B (dense), and Qwen3-30B-A3B (MoE), downloaded from HuggingFace
(Llama-3.2 is gated; set \texttt{HF\_TOKEN}).

\subsection*{Reproducing the results}
{\footnotesize
\begin{verbatim}
git clone --recursive -b tgx-osdi26-ae \
   https://github.com/mirage-project/mirage
cd mirage && bash artifact_evaluation/setup.sh
bash artifact_evaluation/<gpu>/run_tgx.sh
bash artifact_evaluation/<gpu>/run_pytorch.sh
bash artifact_evaluation/<gpu>/run_vllm.sh
bash artifact_evaluation/<gpu>/run_sglang.sh
\end{verbatim}}
\noindent All runs use offline batched inference with a fixed prompt length of 64,
1024 decoded tokens, and greedy decoding; each reported number is the median of
five runs after a four-iteration warmup. On a 40\,GB A100 the sweep covers the four
smaller models; reproducing the Qwen3-30B-A3B point additionally requires an
80\,GB A100 (or an H100/B200).
\end{revblock}

\end{document}